\documentclass[conference]{IEEEtran}
\IEEEoverridecommandlockouts
\usepackage{cite}
\usepackage{amsmath,amssymb,amsfonts}
\usepackage{algorithmic}
\usepackage{graphicx}
\usepackage{textcomp}
\usepackage{hyperref}
\usepackage{xcolor}
\usepackage{ulem}
\usepackage{tcolorbox}
\usepackage{afterpage}

\def\BibTeX{{\rm B\kern-.05em{\sc i\kern-.025em b}\kern-.08em
    T\kern-.1667em\lower.7ex\hbox{E}\kern-.125emX}}

\newcommand{\nip}[2][0pt]{\vspace{#1}\noindent\textbf{#2}}

\begin{document}

\title{Reference Architecture of a Quantum-Centric Supercomputer}

\author{
    \IEEEauthorblockN{Seetharami Seelam, Jerry M.~Chow, Antonio Córcoles, Sarah Sheldon, Tushar Mittal, Abhinav Kandala, \\ Sean Dague, Ian Hincks, Hiroshi Horii, Blake Johnson, Michael Le, Hani Jamjoom, and Jay M.~Gambetta}
    \IEEEauthorblockA{IBM T.J. Watson Research Center, Yorktown Heights, NY 10598, USA\thanks{Corresponding authors:\{sseelam,chowmj\}@us.ibm.com } \\
    }
}
\maketitle

\begin{abstract}
Quantum computers have demonstrated utility in simulating quantum systems beyond brute-force classical approaches. As the community builds on these demonstrations to explore using quantum computing for applied research, algorithms and workflows have emerged that require leveraging both quantum computers and classical high-performance computing (HPC) systems to scale applications, especially in chemistry and materials, beyond what either system can simulate alone. Today, these disparate systems operate in isolation, forcing users to manually orchestrate workloads, coordinate job scheduling, and transfer data between systems -- a cumbersome process that hinders productivity and severely limits rapid algorithmic exploration. These challenges motivate the need for flexible and high-performance Quantum-Centric Supercomputing (QCSC) systems that integrate Quantum Processing Units (QPUs), Graphics Processing Units (GPUs), and Central Processing Units (CPUs) to accelerate discovery of such algorithms across applications. These systems will be co-designed across quantum and classical HPC infrastructure, middleware, and application layers to accelerate the adoption of quantum computing for solving critical computational problems. We envision QCSC evolution through three distinct phases: 
(1) quantum systems as specialized compute offload engines within existing HPC complexes; (2) heterogeneous quantum and classical HPC systems coupled through advanced middleware, enabling seamless execution of hybrid quantum-classical algorithms; and (3) fully co-designed heterogeneous quantum-HPC systems for hybrid computational workflows. This article presents a reference architecture and roadmap for these QCSC systems.
\end{abstract}


\section{Introduction}
Quantum computers have demonstrated utility in simulating quantum systems beyond brute-force classical approaches~\cite{kim2023evidence,roncevic2026mobius}. 
The community has begun developing algorithms and workflows leveraging both 
quantum computers and classical high-performance computing (HPC) systems 
to scale applications, especially in chemistry and materials, beyond what either system can simulate alone~\cite{robledo2024chemistry,roncevic2026mobius}.

This convergence of quantum computing (QC) and high-performance computing (HPC) represents a paradigm shift in computational science, promising to address problems that remain intractable for classical HPC systems alone. As quantum processors advance in scale and accuracy of quantum circuit execution, the integration of Quantum Processing Units (QPUs) with classical HPC infrastructure -- comprising Graphics Processing Units (GPUs) and Central Processing Units (CPUs) -- emerges as a critical pathway toward solving grand challenge problems in materials science, drug discovery, and optimization at unprecedented scales.

Quantum-Centric Supercomputing (QCSC) systems~\cite{qcsc} represent this convergence, where quantum and classical resources are not merely co-located but are deeply integrated through co-designed infrastructure, middleware, and applications. Unlike standalone quantum computers or classical HPC systems operating in isolation, QCSC systems leverage the complementary strengths of each computing paradigm: quantum processors can provide speedups for computational tasks that involve generating and sampling from correlation structures that require exponential resources to represent classically but arise naturally in quantum circuits. Specifically, quantum computing has the potential to demonstrate computational advantages in the following areas:

\subsection*{Hamiltonian Simulation} 

Accurate simulation of quantum many-body systems is widely regarded as one of the most promising applications of quantum computing, with relevance to chemistry, materials science, and condensed-matter physics. A central task in this setting is the estimation of low-energy spectra of many-body Hamiltonians. In the fault-tolerant regime, quantum phase estimation (QPE) is a standard approach for this problem~\cite{kitaev1995quantum}. However, practical implementations of QPE require circuit depths that remain beyond the capabilities of current quantum processors.

These limitations have motivated the development of algorithms with lower circuit-depth and runtime requirements for current devices. Sample-based Quantum Diagonalization (SQD) is one such approach~\cite{robledo2024chemistry}. It uses quantum processors to generate samples from quantum circuits, which are then processed within a QCSC workflow to estimate ground-state energies. Extensions such as Sample-based Krylov Quantum Diagonalization (SKQD)~\cite{yu2025quantum}  and SqDRIFT~\cite{piccinelli2025quantum} produce these samples from structured quantum circuits that define subspaces better suited to ground-state estimation, and in certain settings these approaches admit convergence guarantees.

SQD has already been applied to small molecular systems relevant to drug discovery, while SKQD has been used to study impurity models. SqDRIFT was also recently shown to capture the underlying physics of a newly physically synthesized half-Möbius molecule~\cite{roncevic2026mobius}. Scaling these approaches to larger systems will likely require more advanced quantum-classical workflows. In particular, embedding techniques from computational chemistry decompose a complex system into smaller fragments, allowing the most challenging parts to be treated with higher-accuracy methods while the remainder is handled at lower cost. Within a QCSC framework, the quantum processor can target the hardest fragments, while classical resources perform the preprocessing and post-processing needed to combine the fragment contributions~\cite{shajan2026proteins_SQD}.

\subsection*{Optimization} 

Combinatorial optimization problems lie at the core of many real-world applications, e.g., in finance and logistics. Quantum optimization algorithms such as QAOA~\cite{farhi2014qaoa}, and its extensions to multi-objective optimization~\cite{kotil2025moo}, offer potential advantages for exploring difficult problems. Recent benchmarking studies~\cite{koch2025qoblib} have identified concrete problem classes where quantum heuristics could provide meaningful benefits, while also clarifying where classical baselines remain strong. At the same time, a critical question for practical applications of quantum optimization algorithms is how to efficiently encode the problem into quantum circuits. The typical approach is to formulate the problem as a quadratic unconstrained binary optimization (QUBO) and use QAOA circuits. Ongoing work has shown that extensions such as higher-order binary optimization (HUBO) representations can expose structure that is more amenable to quantum computers and classical-quantum co‑design~\cite{Romero2025hubo, chandarana2025runtimeadvantage}.

These insights motivate hybrid workflows that integrate quantum sampling with classical optimization, including warm‑start strategies~\cite{egger2021warmstart}, and iterative refinement~\cite{bravyi2020variational}. Early demonstrations indicate that quantum samples can guide classical search towards promising regions of the solution space, suggesting a pathway toward scalable workflows on real-world optimization tasks~\cite{sachdeva2026integratedpipeline}.

\subsection*{Quantum machine learning}

Quantum computing may create new opportunities for machine learning by providing access to feature representations and probability distributions that are difficult to construct efficiently using classical methods alone. This has led to growing interest in quantum machine learning, where quantum processors are used to construct feature maps, kernels, and generative models~\cite{Havlicek2019kernels,gao2022generative,huang2025generative,layden2025flowmodels}. In particular, results from quantum kernel methods provide evidence that quantum representations can capture structure relevant to the learning task that is not readily accessible classically~\cite{glick2024kernels}. Meanwhile, recent advances in quantum embedding techniques are expanding the space of representations that can be used in learning tasks~\cite{candelori2026learning}. More broadly, these developments suggest a natural role for hybrid quantum-AI workflows, in which CPUs, GPUs, and QPUs are used together to combine classical optimization and data processing with quantum resources for sampling, feature construction, and representation learning.

\subsection*{Partial differential equations }

Solving differential equations often requires intensive computing resources for classical numerical simulations, making problems like computational fluid dynamics some of the most challenging to compute. Quantum algorithms using HHL to solve discretized PDE/ODE are infeasible to implement today as the quantum circuits are too deep for current quantum computers and the sampling complexity is amplified by bad condition numbers. Recent work develops the first end-to-end quantum algorithms for simulating noisy nonlinear dynamics with polynomial scaling~\cite{bravyi2025quantumsimulationnoisyclassical}. Applications of this algorithm include solving the Navier Stokes equation for simulating turbulent flow. Other work has considered using quantum algorithms for linearizing the Vlasov equation, a nonlinear partial differential equation relevant for plasma physics~\cite{ameri2023vlasov}. Workflows that efficiently combine quantum and classical resources will be necessary for solving these problems at scale. Within the QCSC setting, quantum solvers for differential equations may augment state-of-the-art classical solvers in regime that would otherwise be out of reach.

\subsection{Motivation for Quantum-Centric Supercomputing Systems}

Realizing the full potential of quantum computing systems requires the development of not just quantum system technology but an integrated Quantum-Classical systems architecture. We call this a Quantum Centric Supercomputing systems architecture.  Current quantum systems face significant challenges including limited qubit counts ($10^2-10^3$ qubits), and error rates ($10^{-4}$ to $10^{-2}$ per gate), as well as reduced connectivity or execution speed, depending on the platform. These constraints limit the computational reach of quantum processors to the point where specific techniques -- to deal with noise and circuit limitations -- need to be introduced, namely quantum error mitigation~\cite{Temme_2017,Li_2017} and quantum error correction~\cite{PhysRevA.52.R2493,PhysRevLett.77.793,Terhal_2015}. For example, quantum error mitigation techniques provide a way for us to evaluate accurate expectation values of observables from noisy, shallow depth quantum circuits, and  error detection checks like spacetime codes~\cite{martiel2025lowoverheaderrordetectionspacetime} allow for generation of highly entangled resource states, all before the maturity of fault tolerant quantum error correction. Not only have these protocols proven rather successful in increasing the computational reach of quantum computers~\cite{kim2023evidence}, but they have started to become increasingly amenable to exploit high-performance classical resources, both accelerated and otherwise.

Looking forward, we also anticipate that developing fault-tolerant solutions for large scale quantum computers will heavily involve HPC capabilities throughout, not only at the system level, but also potentially at specific levels for quantum error correcting code definition.

Simultaneously, many computational problems exhibit hybrid characteristics—quantum subroutines that address exponentially scaling components embedded within larger classical workflows that requires HPC systems~\cite{Alexeev_2024,sqd-diag}. These patterns motivate a phased approach to QCSC development.

\subsection{QCSC Evolution: A Three-Phase Roadmap}
\begin{figure*}[h!]
    \centering
    \fbox{\includegraphics[width=0.9\linewidth]{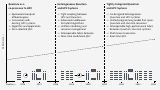}}
    \caption{\bf{Our QCSC roadmap delivers progressively advanced Quantum and Classical HPC systems with seamless stack-level integration to enable high-performance workload execution.}}
    \label{fig:qcscroadmap}
\end{figure*}

We envision QCSC evolution through three distinct phases over the next several years as shown in Figure~\ref{fig:qcscroadmap}:
\begin{enumerate}
    
\item \textbf{Phase 1}: Quantum specialized compute engines - Quantum systems function as specialized compute offload engines within existing HPC complexes, similar to how GPUs augment CPU-based systems~\cite{arie04}. 
This phase focuses on establishing foundational integration across hardware, resource management software and workflow managers, enabling quantum resources to be managed within familiar HPC software tools. As these efforts mature, emerging applications use integrated quantum-HPC systems for closed-loop hybrid computations and error mitigation algorithms, demonstrating quantum offload viability while informing requirements for deeper integration. 

\item \textbf{Phase 2}: Heterogeneous quantum and classical HPC systems - Quantum and classical resources become tightly coupled through advanced middleware, enabling seamless workload distribution, dynamic resource allocation, and sophisticated error mitigation strategies that span both computing domains.

\item \textbf{Phase 3}: Co-designed heterogeneous quantum-HPC systems - Fully co-designed quantum-HPC systems emerge, featuring unified programming models, integrated system software, and hardware architectures optimized for quantum-classical synergy from the ground up. Systems efficiently partition problems between quantum processors (exponential solution spaces) and classical processors (polynomial-time pre-processing and analysis). 
\end{enumerate}

This roadmap is described in detail in Section~\ref{sec:roadmap}.

\subsection{Outline}
This article presents a comprehensive multi-year roadmap for QCSC development, including a reference architecture that addresses key technical challenges across hardware integration, system software, programming abstractions, and application domains. We describe the infrastructure requirements, platform capabilities, and application characteristics necessary at each evolutionary phase. Our goal is to establish a practical framework that guides the quantum computing and HPC communities toward realizing the transformative potential of quantum-centric supercomputing. We begin by describing the application patterns that guide our design.

\subsection{Definition of terms used in QCSC architecture}
Here we define common terms used throughout this paper:
\begin{itemize}
\item \textit{Real-time:} Computational interactions where quantum and classical systems exchange data with minimal latency (typically microseconds) to support tight coupling within a single computational loop. In this mode, classical systems must be custom-designed and programmed to respond within the coherence time of the quantum processor, requiring physical co-location and low-latency interconnects.
\item \textit{Near-time:} Computational interactions where quantum and classical systems exchange data with moderate latency (tens of microseconds to seconds), allowing for iterative workflows where classical processing occurs between quantum circuit executions. 
\item \textit{Batch-time:} Computational interactions where quantum and classical systems operate with relaxed latency requirements (seconds to hours). Both quantum and classical jobs are submitted to their respective job queues, executed independently, and results are communicated via shared storage or standard data transfer mechanisms.
\item \textit{Temporal coupling:} The degree of time-dependent coordination required between quantum and classical tasks. Strong or tight temporal coupling is required when quantum and classical operations must be synchronized within real-time or near-time windows, while weak or loose temporal coupling allows for batch-time execution with flexible scheduling.  
\item \textit{Spatial coupling:} The degree of physical proximity and interconnection bandwidth required between quantum and classical computing resources. Strong or tight spatial coupling implies co-location within the same facility or rack with high-bandwidth or low-latency or both, whereas weak or loose spatial coupling allows for geographically distributed resources connected via standard network infrastructure. Strong temporal coupling typically needs tight spatial coupling. 
\end{itemize}

\section{QCSC application patterns that guide the reference architecture}

To ground our discussion of the requirements for the reference architecture, we begin with representative hybrid use cases that illustrate the diverse temporal and spatial coupling requirements arising in practical QCSC applications. 

\subsection{Use Case 1: Electronic Structure Calculations}

Robledo-Moreno et al.~\cite{robledo2024chemistry} and Kanno et al.~\cite{kanno2023QSCI} developed a Sample-based Quantum Diagonalization (SQD), a hybrid quantum-HPC method that enables  electronic structures calculations beyond the computational limits of exact diagonalization. 
In this case, the quantum processor is treated as a specialized offload engine whose role is restricted to an \textit{intrinsically quantum} portion of the workflow. In particular, executing a limited number of large circuits and returning samples, while all other steps of the workflow are distributed across classical HPC nodes.  

Concretely, the system architecture features loose spatial coupling and weak temporal coupling between heterogeneous quantum and HPC systems. The quantum system can either be co-located with the classical HPC system or connected over the cloud, and both systems are managed by independent workload managers operating in batch-time mode.  

In the SQD workflow, the chemistry Hamiltonian is first mapped to qubits, then the QPU generates bitstring samples representing electronic configurations (Slater determinants). 

The classical HPC supercomputer performs the following operations on the samples (see Figure 1 in~\cite{robledo2024chemistry}):
\begin{itemize}
    \item Partitions the samples into two groups based on hamming weight:
    \begin{enumerate}
        \item \textit{Correct hamming weight}: Samples with the expected number of electrons proceed unchanged to the next step
        \item \textit{Incorrect hamming weight}: Samples with incorrect hamming weights due to noise undergo configuration recovery to restore the correct hamming weight matching the number of electrons (recovered samples)
    \end{enumerate}
    \item Combines the correct hamming weight samples with the recovered samples
    \item Performs subsampling on the combined set
    \item Executes subspace projection and diagonalization of the Hamiltonian in the selected determinant basis (the most computationally intensive step)
\end{itemize}
 
A key architectural point, shown in that figure, is that the SQD process on the classical HPC is executed in a self-consistent loop among these three steps: the diagonalization outputs update a reference occupancy vector, which is fed back into the configuration recovery and the loop repeats until convergence or exceeding a certain number of iterations. Here, the QPU provides the sampling signal and the HPC system iterates and refines the sparse wave function approximation as shown in Figure~\ref{fig:sqd}(a). 

Robledo-Moreno et al. demonstrated this approach using an IBM Heron Quantum processor and RIKEN's Fugaku supercomputer~\cite{fugaku_supercomputer} to simulate ground-state energies of $N_2$, $[2Fe-2S]$, and $[4Fe-4S]$ clusters, with circuits up to 77 qubits and 10,570 gates. They report the largest run using 6,400 Fugaku nodes out of the ~152,000 nodes of the full system. 

In this design, orchestration between quantum and classical is sequential and only requires loose coupling between the two systems. The classical system executes its pre-processing tasks based on the scheduling policies, job priorities, resource requirements and when the pre-processing is done, a job is submitted to the quantum system, which executes the job to collect the samples based on its own scheduling policies and job priorities. Once the samples are available, another job is submitted on the classical HPC system using only a small number of nodes (~6,400) to performance the large-scale diagonalization in a self-consistent loop. 

This kind of workflow can be supported by having systems co-located or systems in distinct locations with HPC systems on-prem and Quantum in the cloud, or both in the cloud or Quantum on-prem and HPC in the cloud.

However, this approach faces a key challenge: what if the workflow requires iterative execution between quantum and classical systems? Specifically, what if classical computation results could be used to refine quantum circuit parameters and generate improved bitstrings? This iterative feedback scenario is addressed in the next use case. 

\begin{tcolorbox}
Key Observation 1: SQD exemplifies batch-time integration with loose coupling between quantum and classical HPC systems. These workflows do not require spatial or temporal coupling or co-location of systems, enabling flexible deployment across distributed or cloud connected infrastructures.
\end{tcolorbox}

\afterpage{
\begin{figure*}[h!]
    \centering
   \fbox{\includegraphics[trim=2cm 4cm 2cm 4cm, clip,width=0.95\linewidth]{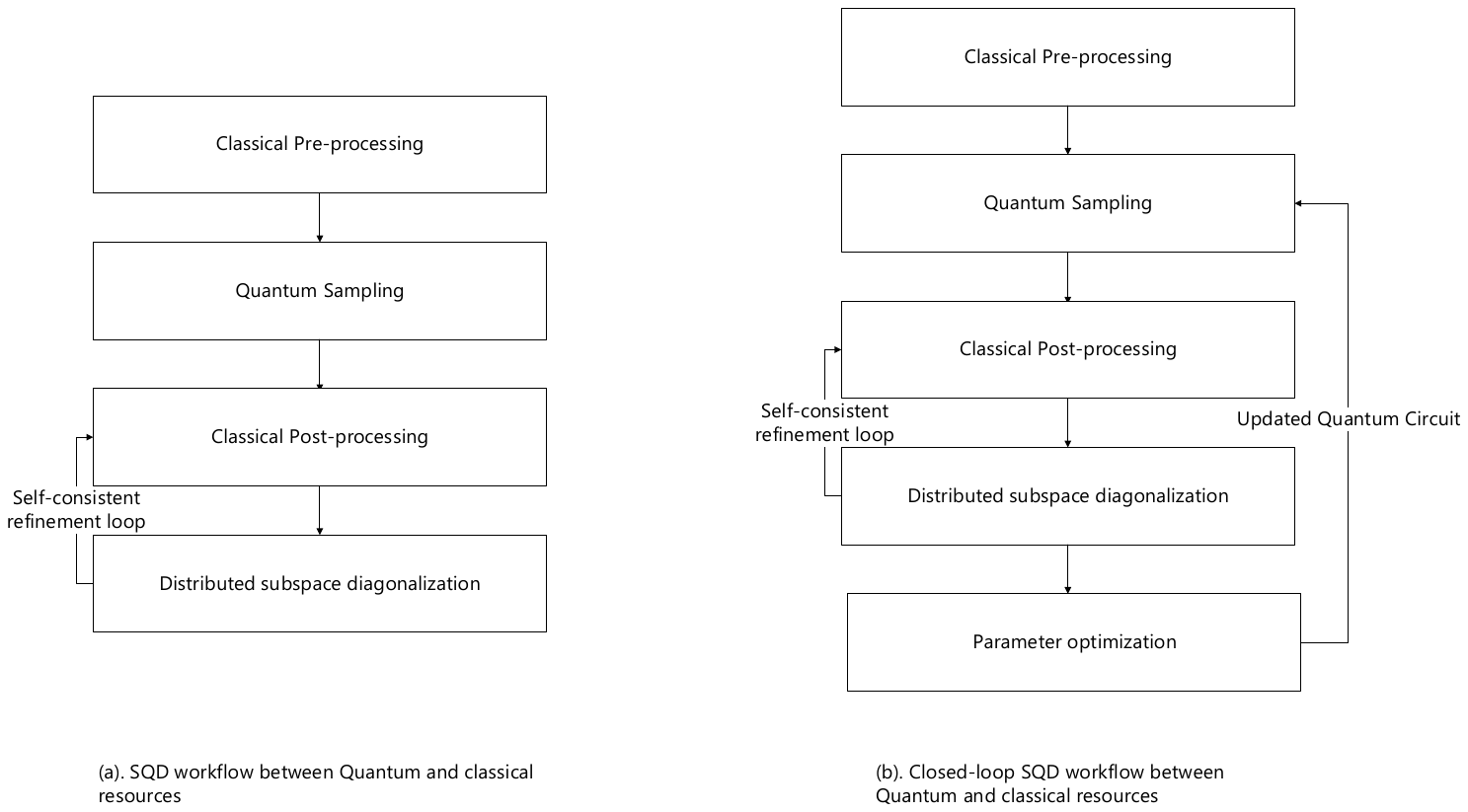}}
    %\vspace{-1cm}
    \caption{\bf{Illustration of the (a). SQD workflow and (b). Closed-loop SQD workflow between Quantum and classical resources. In (a). the results of the classical post processing are used to compute ground-state energies, where as in (b). the results of the classical step are fed to optimize the quantum circuits in subsequent steps to achieve optimal ground-state energies.}}
    \label{fig:sqd}
\end{figure*}
}

\subsection{Use Case 2: Closed-Loop Electronic Structure Calculations}
Shirakawa et al.~\cite{shirakawa2025closedloopcalculationselectronicstructure} demonstrated closed-loop electronic structure calculations using an on-premises IBM Heron Quantum system along with the Fugaku supercomputer. This work introduces a more tightly integrated orchestration model that treats the quantum-classical workflow as a co-scheduled distributed optimization problem. Instead of simply iterating an estimator loop as in use case 1, this method iteratively refines a quantum state representation of the molecular wavefunction using quantum sampling, classical subspace diagonalization, and parameter optimization as shown in Figure~\ref{fig:sqd}(b). The loop continues until the ground-state energy converges. This architecture enables electronic structure calculations beyond the reach of traditional classical full configuration interaction methods (e.g., CCSD, HF) by combining quantum sampling efficiency with massively parallel classical computations. 

In this case, half of the entire Fugaku system with over 72,000 nodes and the full Fugaku system with over 150,000 nodes are used concurrently with their IBM Heron based quantum system that is also on-premises. 

A key characteristic of this workflow is the tight temporal and spatial coupling between quantum hardware and the classical HPC system. This coupling is required for a few reasons: (1) continuous data exchanges between the classical system and the quantum system as each completes their step, the output is used in the following step on the other system, (2) because the quantum and classical computations are in a loop where both system resources are essentially dedicated to this workflow, the data must be transferred in a timely manner from a latency and bandwidth perspective, and (3) It is possible to dedicate both systems for this workflow simply because both systems are co-located and managed by a single organization that has control on the job queues and user priorities on these systems. If either or both of these systems are accessed via cloud, and shared with a large group of users from a different organization, dedicating time across both systems will be a significantly challenging task. For these reasons, a temporally and spatially co-located Quantum and Classical HPC system is desired to orchestrate such closed loop computations. 

\begin{tcolorbox}
Key Observation 2: Closed-loop SQD exemplifies batch-time integration with temporal and spatial coupling.  While the classical and quantum systems executed as distinct jobs, the output from one system (e.g., classical result) is used to refine the parameters for the next iteration on the other system (e.g., quantum circuit parameters), creating an iterative feedback loop. Spatial and temporal co-location of these systems enables efficient execution of such closed loop computations.
\end{tcolorbox}

\subsection{Use Case 3: Embedding Classical Processing in the Quantum Workflow}

Classical computation can extend the reach of pre–fault-tolerant quantum processors, and vice versa. Several examples of this interplay arise in the context of error mitigation techniques~\cite{temme2017error,li2017efficient}. In contrast to quantum error correction, error mitigation trades qubit overhead for sampling or runtime overhead, estimating noise-free expectation values by executing ensembles of noisy quantum circuits. Despite the exponential scaling of this sampling cost, these methods have enabled accurate observable estimation on circuits with $\sim$100 qubits and a few thousand two-qubit gates~\cite{kim2023evidence}, providing an immediately accessible path toward quantum advantage~\cite{lanes2025framework}.

Recent developments integrating classical HPC with error-mitigation workflows offer a path to reducing this sampling overhead, in exchange for additional classical runtime and, in some cases, controlled bias. One example is tensor-network error mitigation (TEM)~\cite{filippov2023scalable,fischer2026dynamical}, which employs tensor-network calculations to effectively invert the noise. In this approach, the task of estimating the expectation value of an observable in the ideal circuit is mapped onto estimating a modified observable on the noisy quantum circuit. TEM employs a middle-out tensor contraction of a near-mirror version of the circuit and is expected to be more efficient than tensor-network simulation of the original circuit itself.

These hybrid workflows also benefit from recent advances in classical simulation techniques. In particular, Pauli propagation methods~\cite{rall2019pauliprop,begusic2025cpt,rudolph2025pauliprop,fuller2025obp} have emerged as a powerful classical simulation techniques for observable estimation, producing competitive results relative to both, other classical heuristics and quantum simulation experiments. Pauli-propagation methods can quantify how Pauli errors occurring at different locations in the circuit affect a target observable, defining a shaded light cone of observable sensitivity. Such light-cone structures highlight regions of the circuit where error-mitigation resources can be concentrated to most efficiently cancel the noise, leading to order-of-magnitude runtime reductions~\cite{eddins2024lightcone}.

Traditional Pauli propagation relies on back-propagating the observable through the entire circuit, with truncations introduced to manage the exponential growth in operator paths. Hybrid Pauli-propagation workflows that incorporate executions on a noisy quantum processor can estimate the contributions of truncated paths, achieving higher accuracy with lower classical resources than purely classical approaches~\cite{QuEPP2026}. Alternatively, on sufficiently low-error quantum processors, it can be more efficient to forward-propagate near-identity Pauli noise channels and measure a modified observable, rather than performing full operator back-propagation~\cite{PNA2026}.

Beyond circuit noise, tensor-network methods can also mitigate algorithmic errors. For example, tensor-network calculations combined with multi-product formulas~\cite{zhuk2024trotter,robertson2025tensor} can suppress Trotterization error in digital time-evolution simulations, extending the reach of near-term quantum processors with limited gate budgets.

These emerging error-mitigation workflows place increasing demands on classical computing resources. Tensor-network contractions, Pauli-propagation analysis, and large-scale sampling require substantial CPU and GPU capability, high-throughput data movement, and efficient orchestration between quantum executions and classical post-processing. As circuits grow in width and depth, the computational effort associated with identifying light-cone structures, contracting tensor networks, or optimizing mitigation strategies can rival or exceed the cost of the quantum experiments themselves. Consequently, effective deployment of these techniques benefits from access to HPC infrastructure capable of performing large-scale classical analysis alongside quantum workloads. Integrating quantum processors with HPC resources enables these mitigation pipelines to operate efficiently, allowing classical simulation, data reduction, and mitigation optimization to proceed in parallel with quantum execution and thereby extending the practical reach of pre–fault-tolerant quantum systems.

\begin{tcolorbox}
Key Observation 3: Advanced error-mitigation techniques shift much of the computational burden to classical resources. Tensor-network contractions, Pauli propagation, and large sampling campaigns require substantial CPU/GPU compute, high-throughput data movement, and rapid orchestration between quantum execution and classical analysis. As quantum circuits scale, this access to HPC infrastructure becomes a key enabler for extending the reach of QPUs. 
\end{tcolorbox}

\subsection{Use Case 4: Open Loop Error Correction Research}
Quantum error-correction (QEC) research and Development (R\&D) is a systems-level discipline spanning hardware, control, coding theory, and classical compute. Modern QEC experiments increasingly rely on large-scale dynamic circuits with mid-circuit measurement and conditional logic, such as in the exploration of magic state production~\cite{Gupta_2024, brown2023advancescompilationquantumhardware}. In addition, recent advances in spacetime checks~\cite{martiel2025lowoverheaderrordetectionspacetime, dasu2026computingencodedlogicalqubits, Liao_2025, reichardt2025faulttolerantquantumcomputationneutral} demonstrate that error detection need not be confined to static stabilizer rounds, but can instead be constructed across extended space–time regions of a dynamic circuit, enabling structured detection of correlated errors in regimes previously inaccessible to practical experimentation. These workloads demand deterministic timing, high-bandwidth measurement capture, and tight integration between the QPU and classical resources. 

These advances in error detection --- such as the spacetime checks applied across large dynamic circuits --- do more than suppress errors; they expand the frontier of quantum states that can be reliably prepared and interrogated~\cite{martiel2025lowoverheaderrordetectionspacetime}. By pushing detection capability deeper into complex circuit regions, researchers can stabilize and explore increasingly intricate entangled states that would otherwise decohere before meaningful study. This, in turn, opens new directions in algorithm research: detection-enabled state preparation, deeper variational circuits with structured filtering, and hybrid detection-plus-mitigation workflows. 

Beyond error detection, we observe that advances in QEC research can meaningfully be performed without decoding in a strict real-time feedback loop. A substantial portion of innovation --- decoder development, hierarchical partitioning strategies, code comparisons, and threshold estimation --- can be conducted through offline decoding workflows. In this mode, measurement streams are passed through high-bandwidth links to scale-out GPUs, TPUs~\cite{jouppi2017indatacenterperformanceanalysistensor}, CPUs or larger HPC resources without real-time feedback. This direction for QEC research opens up a diversity of approaches that are being pursued in the community, such as AI-based decoders (e.g. AlphaQubit~\cite{Bausch2024HighAccuracyDecoding}), to attempts to use soft information~\cite{pattison2021improvedquantumerrorcorrection}, to potentially mixture-of-models approaches which re-analyze the same data set with multiple decoders or decoder configurations. Some of these methods will leverage increased data rates between the QPU and the classical compute resources. This research benefits from rapid iteration on decoder algorithms without being constrained by microsecond latency budgets, while still also informing towards future low-latency integration.

\begin{tcolorbox}
Key Observation 4: Quantum error correction and detection research can be performed with flexible classical processors operating with quantum execution over high-bandwidth links. 
\end{tcolorbox}

\begin{figure}[h!]
    \centering
    \includegraphics[trim=2cm 2cm 2cm 2cm, clip,width=1.0\linewidth]{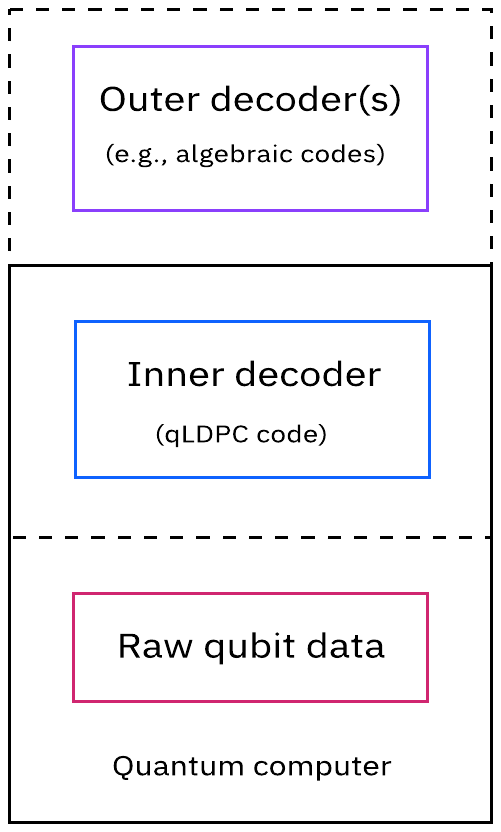}
    \vspace{-1cm}
    \caption{\bf{Hierarchical error correction for future fault-tolerant workflows. Hierarchical codes involve two levels of coding: (1) initial error correction of raw qubit data and (2) additional noise suppression
}}
    \label{fig:innerouter}
\end{figure}

\afterpage{
\begin{table*}[h!]
\centering
\caption{Mapping of use cases to QPU--HPC integration modes.}\
\label{tab:integration_modes}
\begin{tabular}{|l|l|l|l|l|}\hline
\textbf{Use Case} & \textbf{Temporal Coupling} & \textbf{Spatial Coupling} & \textbf{Classical Resource} & \textbf{Key Requirement} \\\hline
1. Electronic Structure (SQD) 
& Batch-time 
& Loose 
&  HPC Supercomputers
& Co-location optional \\\hline

2. Electronic Structure (Closed-loop SQD)
& Batch-time 
& Tight 
& HPC Supercomputers 
& Co-location preferred \\\hline

3. Error Mitigation 
& Batch-time/Near-time (future)
& Tight  
& CPU and GPU HPC systems
& High bandwidth \\\hline

4. Open Loop QEC R\&D 
& Batch-time 
& Loose 
& CPU, GPU, TPU HPC systems
& High bandwidth \\\hline

5. QEC outer decoder R\&D 
& Near-time 
& Tight 
& CPU, GPU, TPU HPC systems
& Low-latency \\\hline
\end{tabular}
\end{table*}
}

\subsection{Use Case 5: Outer code research for fault-tolerant Quantum Error Correction}

Fault-tolerant quantum computing will likely rely on hierarchical error correction, in which multiple layers of coding operate at different physical and algorithmic timescales. Rather than relying on a single monolithic code, this approach separates the problem into an inner hardware-adapted code that stabilizes physical qubits in real time and an outer algorithm-adapted code that provides additional logical protection and noise suppression~\ref{fig:innerouter}. The concept builds on the original theory of concatenated quantum codes, where logical qubits are recursively encoded to reduce logical error rates below threshold. Modern developments in quantum error correction, including recent advances in quantum low-density parity check (LDPC) codes and scalable fault-tolerant architectures further motivate layered approaches that align each level of coding with the latency and computational capabilities of surrounding classical control and compute systems.

The inner code operates directly on physical qubits and must respond on hardware timescales. Syndrome measurements are generated every few microseconds, requiring decoding and feedback within $\sim10\,\mu$s control loops tightly integrated with highly customized QPU control electronics such as ASICs and FPGAs and leveraging sub-microsecond latency interconnects for superconducting qubit-based QPUs. This regime is well suited to hardware-adapted quantum LDPC (qLDPC) codes, whose sparse parity-check structure enables scalable error correction while remaining compatible with processor connectivity and control constraints. Recent advances in qLDPC constructions~\cite{10.5555/2685179.2685184, PhysRevLett.129.050504,10.1145/3519935.3520017,10103665} together with IBM architectural proposals such as the Gross code framework and related implementations~\cite{Bravyi2024LDPC,Yoder2025TourDeGross}, highlight how such codes can form an efficient hardware-level protection layer capable of stabilizing large arrays of physical qubits. Since this real-time inner code QEC is inherently integrated within the quantum system architecture, we do not consider it a candidate workload for QCSC.

However, an outer code can then operate on the logical qubits produced by the inner layer, where syndrome information is generated at a slower rate—typically $\sim100\,\mu$s to $1\,\mathrm{ms}$. In a quantum-centric supercomputing architecture, this layer naturally maps to nearby scale-up GPU or other classical accelerator nodes, and provides a natural environment for continued research and exploration in quantum error correction, allowing new decoding algorithms, logical code constructions, and hybrid quantum-classical inference techniques to be developed and evaluated without disrupting the tight real-time feedback loops required by the underlying hardware. In this hierarchical structure, the inner layer rapidly stabilizes raw qubit data, while the outer layer performs deeper inference and additional noise suppression across larger logical structures.

\begin{tcolorbox}
Key Observation 5: Low latency (microseconds) scale-up classical systems of CPUs and GPUs have the potential to enable outer code research towards future fully-integrated fault-tolerant quantum error corrected systems, that inherently perform hardware-efficient real-time inner fault-tolerant error correction codes, such as qLPDC codes. 
\end{tcolorbox}

\subsection{Summary of the workload patterns for QCSC}

The key observations described above present different integration patterns between Quantum and Classical computing systems, differentiated by their coupling requirements, latency constraints, classical resource requirements, data sovereignty, and security requirements. 
Table~\ref{tab:integration_modes} summarizes these requirements.

As shown in the table, the first pattern involves batch-time integration with loose coupling, where systems operate independently without spatial or temporal co-location requirements exemplified by standard SQD workflows where quantum and classical jobs run separately. This is the most common pattern among workloads running on IBM Quantum Platform today where customers use IBM Quantum Platform via Cloud API (IQP) and use their on-premises or cloud HPC resources to run SQD-type workloads (see e.g., ~\cite{robledo2024chemistry, Alexeev_2024, smith2025quantumcentricsimulationhydrogenabstraction_SQD,Liepuoniute2025_SQD, shajan2026proteins_SQD, Barison_2025_ext-SQD,kirby2026observationimprovedaccuracyclassical}).

The second pattern represents batch-time integration with tight spatial coupling through closed-loop workflows, where outputs from one system inform the next iteration on the other, such as in closed-loop SQD with iterative parameter optimization that benefits from co-location for efficiency while maintaining a batch execution model. While it is technically feasible to orchestrate such close loop execution between a supercomputer system and IBM Quantum Platform via Cloud API, it requires substantial coordination. The systems available via IQP are accessed by many organizations and users with several jobs in the queue, which makes it hard to pause all the jobs to prioritize these closed-loop jobs. On the other hand, when the Quantum system is co-located with HPC system, managed by the same organization, they can prioritize these closed-loop jobs on both HPC and the Quantum system so they can execute concurrently as described in this paper~\cite{shirakawa2025closedloopcalculationselectronicstructure}. We are seeing an increased use for this pattern of execution in our customer engagements. This not only ensures timely execution of these coupled Quantum-Classical jobs, but it also ensures that all of the data associated with the job stays within the boundaries of the organization where these systems are co-located, addressing a key data provenance and security issue. 

The third pattern introduces error mitigation which estimates noise-free quantum results by running multiple noisy
circuits on the quantum system and then using classical
resources to run tensor network methods or Pauli propagation
methods to invert the noise effects in iterative fashion. Today, this is already being explored in batch-time modes, but there is the potential to explore real-time orchestration, high throughput and low-latency connections with this pattern to keep quantum processors from idling during classical post-processing.

The fourth pattern is similar to the first pattern but specifically focuses on error correction and error detection research, where high-bandwidth communication and specialized compute resources such as GPU accelerators and TPUs at scale are essential for large-scale studies.  

The fifth pattern, similar to embedding methods but focused on hierarchical quantum error correction, also requires concurrent classical-quantum operation with real-time feedback loops implemented through co-located accelerators and low-latency interconnects for outer code explorations. 

The key distinction across these modes is a progression from weak spatial and temporal coupling with independent batch jobs, to strong temporal coupling with iterative batch loops, and finally to strong spatial and temporal coupling with concurrent execution and near-time latency requirements.

\section{QCSC Reference Architecture}

\begin{figure*}
    \centering
    \includegraphics[width=0.95\linewidth]{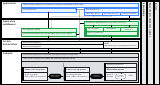}
    \caption{\bf{Quantum Centric Supercomputing Architecture}: the architecture describes how quantum systems can be integrated with high-performance computing infrastructure (CPUs + GPUs) and Heterogeneous Specialized Accelerators in a composable compute layer, with system orchestration, application middleware, and applications with domain-specific libraries built on top.}
    \label{fig:arch}
\end{figure*}

Our QCSC reference architecture defines the full system stack within which Quantum and HPC systems integration occurs.
Figure~\ref{fig:arch} presents this architecture into four horizontal layers from bottom to top: (1) Hardware Infrastructure, (2) System Orchestration, (3) Application Middleware, and (4) Applications, along with cross-cutting layers of (1) Cloud Software, (2) System Management and Monitoring, and (3) Security.

\subsection{QCSC Hardware Infrastructure Layer}
The \textit{Hardware Infrastructure} layer is where the QPU-HPC system integrations are realized in space and time. The figure shows three distinct classical hardware tiers, each with different compute capabilities, proximity, and interconnect relationships to the QPUs. 

The \textit{Quantum System} at the bottom right of Figure~\ref{fig:arch} itself comprises a Classical Runtime and one or more QPUs connected internally by a \textit{real-time} interconnect. The classical system and its runtime in this innermost tier is realized as a heterogeneous ensemble of specialized classical processors including FPGAs, custom ASICs, CPUs, and other purpose-built compute elements, developed and composed to meet the stringent determinism and latency requirements of QPU operations. This is the domain of \textit{tight coupling}, where the classical runtime participates directly in the QPU's operational cycle, performing tasks such as quantum error correction (QEC) decoding, mid-circuit measurements, qubit calibrations, and active qubit reset within the latency budget imposed by qubit coherence times. We consider these technologies to be specific to a Quantum system implementation, and in our case we expect the technologies within to evolve to support the unique needs of our systems as we build towards a fully fault-tolerant system along our roadmap~\cite{ibmquantumroadmap}. 

The capabilities of the Quantum System will be exposed to the rest of the stack through a well-defined \textit{Quantum Systems API (QSA)}, which servers as the sole programmatic boundary between the tightly coupled real-time layer and the rest of the system architecture. The QSA abstracts the internal heterogeneity of the Classical Runtime and the QPU hardware, presenting a stable, potentially vendor-portable interface through which the orchestration layer -- via-QRMI (described later), and the co-located Scale-up tier can be addressed: submit circuits, retrieve results, and manage device configuration without direct knowledge of the underlying classical resources (CPUs, FPGA, ASIC), or quantum resource (QPU) details.  

The \textit{Partner Scale-up Co-located Systems}, in the middle, are CPU, GPU, and Accelerator systems from the likes of Intel, AMD, or NVIDIA that are physically in proximity to the Quantum Systems and connected via a low-latency interconnect such as RDMA over Converged Ethernet (ROCE)~\cite{ibta_roce_2025} , Ultra Ethernet~\cite{hoefler2025ultraethernetsdesignprinciples}, or NVQLink~\cite{nvqlink}. We call this a \textit{near-time} interconnect because of the expected proximity between this scale-up system and the Quantum system as well as the low-latency interconnect that connects these systems. Next-generation scale-up systems will feature heterogeneous specialized accelerators (simply Accelerators) purpose-built for emerging artificial intelligence (AI) workloads~\cite{ruchir}. Here, we anticipate workloads such as in Use Case 5, with the potential for research into new decoding algorithms for outer codes within a hierarchical fault-tolerant error correction framework. 

To enable this a user needs access to a programmable compute resource (e.g., shown in Figure~\ref{fig:decoder}). The co-located CPU, GPU, and Accelerator nodes are well-suited for this task such as training and evaluating decoders against live or recorded syndrome streams for the QPUs. 
\afterpage{
\begin{figure}[h!]
    \centering
    \fbox{\includegraphics[trim=0cm 4cm 0cm 4cm, clip,width=0.9\linewidth]{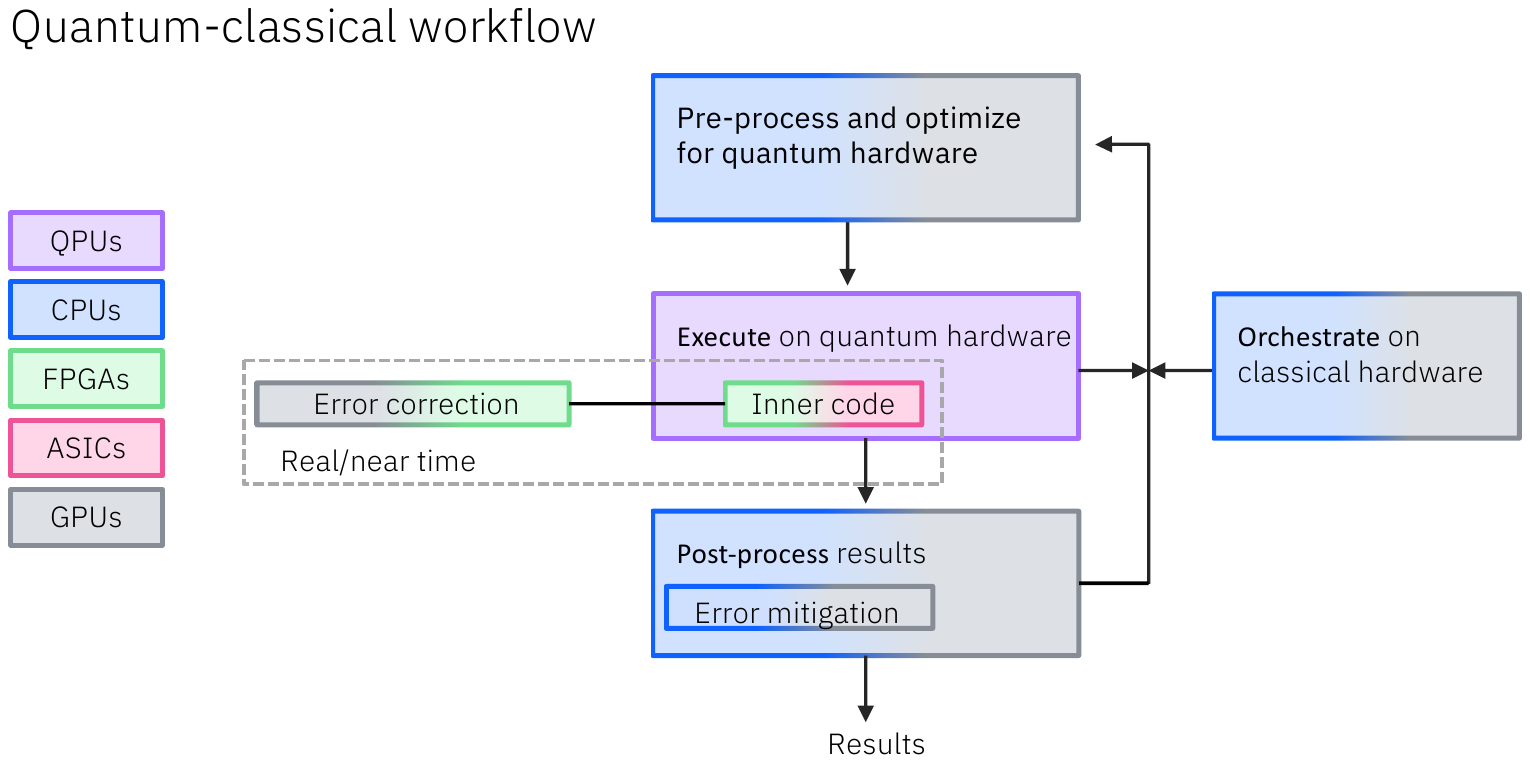}}
    %\vspace{-1cm}
    \caption{\bf{Workflow and orchestration between Quantum and Classical Systems. As indicated by the color legend, GPUs and FPGAs that will be part of the Scale-up systems of Figure~\ref{fig:arch} to support research on outer codes for hierarchical error correction.}}
    \label{fig:decoder}
\end{figure}
}

The \textit{Partner Scale-out Systems} (CPUs+GPUs+Accelerators, Cloud or On-Premises) are connected to the combined Scale-up co-located systems and the Quantum Systems via an appropriately sized high-bandwidth interconnect and represent the broadest and the most flexible classical compute resources in the QCSC stack. These systems are not constrained to a single hardware configuration: they may be deployed as pure GPU clusters optimized for massively parallel HPC and AI workloads, or as mixed GPU, Accelerator, and CPU clusters to support traditional modeling and simulation workloads, and Quantum workloads. This flexibility allows the Quantum systems to be deployed in existing HPC data centers in the near-term, or planned AI factories being scaled out for the future. This configurability also allows the Scale-out tier to be tailored to the specific classical computation profile of the quantum application domain it serves. The Scale-out tier handles the full range of classical workloads that surround and complement QPU execution, including:
\begin{itemize}
    \item \textbf{Classical pre-processing}: 
    This step includes generating initial states of quantum circuits, constructing Hamiltonians, performing orbital optimization, transpiling and optimizing the circuits and sending the quantum circuits to the quantum computer (e.g., see section 3.d in ~\cite{Alexeev_2024}). In algorithms such as SQD (described above)~\cite{shirakawa2025closedloopcalculationselectronicstructure}, the scale-out tier is responsible for constructing the subspace of configurations sampled by the QPU and preparing the inputs for the subsequent diagonalization step.
    \item \textbf{Classical post-processing}:
    This step involves retrieving the results (measurements) from the quantum system, subsampling them, applying error mitigation methods~\cite{Bravyi2021MeasurementMitigation} including recovering configurations~\cite{shirakawa2025closedloopcalculationselectronicstructure}, and calculating observables. For example, in SQD~\cite{shirakawa2025closedloopcalculationselectronicstructure}, this tier is receives serialized results, extracts bitstrings, subsamples them, and recovers configurations, and calculates the energy of the molecule based on the processed measurements using a method called diagonalization where thousands of CPUs cores~\cite{shirakawa2025closedloopcalculationselectronicstructure} and hundreds of GPUs~\cite{sqd-diag} could be used for this work. Note that depending on scale and needs, some of this classical post-processing can be embedded inside a Tensor Compute Graph (see Section \ref{sec:middleware})
    \item \textbf{Classical simulation steps within hybrid workflow}:
    Many quantum algorithms interleave QPU execution with classical simulation substeps. Tensor network methods, density matrix renormalization group (DMRG) calculations, coupled-cluster calculations, and Monte Carlos sampling are examples of simulation workloads that run on the Scale-out tier in coordination with QPU execution, either to provide reference data, to provide a starting point for the quantum circuits, to refine the quantum outputs, or to carry over results from one circuit execution to the next to improve accuracy. Classical HPC is used to simulate small instances or approximate solutions that help validate quantum results and verify correctness. 
\end{itemize}

In QCSC, the interconnect must evolve in tandem with the systems it couples. Our early deployments at RIKEN and RPI utilize Ethernet to connect the scale-out HPC systems with the quantum systems, primarily for handling control traffic, job submission, telemetry, and other batch-oriented tasks. Our architecture is advancing to incorporate co-located scale-up systems featuring low-latency interconnects designed to support error correction research workloads described earlier. While the initial deployments of this architecture may still leverage Ethernet, we anticipate transitioning to RDMA over Ethernet and more advanced protocols in future deployments, as we will describe in the roadmap section. We also expect a high-bandwidth network fabric connecting our scale-up and quantum systems and our scale-out systems, which will continue to evolve over time. To execute workloads seamlessly across these three system boundaries, we recognize a critical need for an interoperable fabric that unifies these heterogeneous systems. 

In addition to networking, we also anticipate the need for an integrated and interoperable storage fabric to support workloads in QCSC. Currently, our quantum systems primarily utilize an object storage interface to send quantum workloads to the system, and return results directly to the clients.  In contrast, the HPC systems rely on high-performance file systems to support workload execution. We envision a convergence of these two storage paradigms to flexibly support the diverse use cases described earlier. Additionally, we anticipate innovations in caching strategies and data representations to enable the tightly coupled workloads expected to run on future QCSC systems.

\subsection{QCSC System Orchestration Layer}

At the \textit{System Orchestration} layer, resources are allocated and coordinated towards the execution of the workflow, which in the most general case will contain interdependent classical and quantum workloads. The exposure of quantum resources from the available system(s) is enabled by a lightweight library, the Quantum Resource Management Interface (QRMI)~\cite{qrmi}, that offers a minimal API to be used by the resource manager (e.g., Slurm~\cite{slurm}) for resource allocation, job scheduling, and job management, including submission, monitoring, cancellation, and execution.  

The overarching goal of the QRMI library is to enable a unified resource management system for efficient coordination of quantum and classical resources while abstracting away hardware-specific details of the system. It stems from the lack of native QPU support in traditional workload managers and the difficulty to coordinate across heterogeneous distributed systems. This shortcoming is clearly evident even in Slurm, one of the most widely deployed systems for orchestrating large-scale classical compute resources. Slurm provides fault‑tolerant, highly scalable job scheduling across many of the TOP500 supercomputers~\cite{top500web}. Its core responsibilities include allocating exclusive or shared access to nodes, launching and monitoring jobs, and enforcing queuing and policy constraints. A key reason for its broad adoption is its modular architecture: Slurm exposes a rich plugin interface that allows sites to extend functionality, such as accounting, custom resource scheduling, topology-aware placement, or license management, without modifying the scheduler's core. However, despite this extensibility, Slurm has no native concept of quantum resources, which motivates the need for frameworks like QRMI to extend its capabilities towards integrating QPUs into its model.

In the specific example of Slurm, the current implementation of QRMI is realized through a quantum SPANK (Slurm Plug-in Architecture for Node and job Kontrol) plugin~\cite{qiskit-slurm}. This plugin exposes the QPUs as generic resources to Slurm and gives Slurm access to the tools for accessibility check and acquisition. It provides these functionalities by invoking the QRMI functions that call the quantum resource directly via, for example, its REST APIs. 

QRMI can also provide interfaces compatible with the backend model of different quantum vendors. That way, the integration with existing quantum software development toolsets like Qiskit~\cite{qiskit} becomes straightforward.

Workflow management tools offer additional benefits beyond simply scripting the workflow. They offer an explicit execution flow, efficient recovery from failures, more interdependency context in classical-quantum hybrid workflows, and enhanced reproducibility.  Workflow managers typically run workflow servers in a login node of a scheduler like for example Slurm. They can also launch executables within specific environments, like MPI or Qiskit.

While the workflow management ecosystem is considerably more fragmented than the resource management ecosystem -- which has converged around widely adopted implementations like Slurm -- extensions to support hybrid quantum-classical workflows have begun to emerge~\cite{qiskit-prefect, tierkreis}.  A critical aspect of these tools is to ensure the efficient use of scarce resources like QPUs in a multi-tenancy context. Several capabilities play instrumental roles in making workflow management both feasible and impactful for hybrid quantum-classical workflows: (1) algorithmic approaches that parallelize quantum and classical computations, either naturally or by design; (2) solutions that enable more accurate runtime estimation for different resources, accounting for quantum-specific factors such as repeat-until-success approaches or sampling overhead estimations based on accurate noise modeling; (3) comprehensive job profiling capabilities that can be used to optimize resource allocation across heterogeneous computing environments. 

The orchestration layer must also accommodate differences between HPC and quantum user‑management models. Slurm uses site‑managed POSIX identities, while quantum systems may require separate credentials or access policies. QRMI can address this mismatch by mediating provider‑level credentials and environment setup at orchestration time, allowing users to request QPUs through Slurm while access control remains encapsulated.

In addition, some deployments may incorporate quantum cloud bursting, where on‑premise quantum hardware is supplemented with externally hosted quantum devices. By presenting these devices as quantum resources and managing acquisition and environment setup through QRMI and the SPANK plugin, the orchestration layer can support cloud bursting transparently and preserve a unified user experience.

\subsection{QCSC Application Middleware}\label{sec:middleware}
The inherently hybrid nature of QCSC calls for very different data constructs to express the different types of complexity found in a computational task. These are captured in the Application middleware layer of~Fig.\ref{fig:arch} , where parallel programming models exist both for advanced classical computations and for quantum computations. 

Traditional HPC classical programming models deal with tensor data structures leveraged by hardware accelerators to store and manipulate information efficiently and exploited by programming models aimed at heavy parallelism. These programming models handle inter-dependencies among tensor data structures, for example in the form of graphs, and define the paradigm for code structure, optimization, and execution.

In the case of quantum, the programming model relies on the computational construct of the quantum circuit. Quantum circuits are ordered sequences of operations, unitary and otherwise, applied to qubits. Quantum circuits can contain classical control flow within them and typically involve measurements, which project the qubits in one of two possible states in their measurement basis. 

Quantum algorithms describe computational problems in terms of quantum circuits. This description can happen at many different levels of abstraction, from fairly high level involving only sets of Pauli rotations, general unitaries, and different mathematical operators, to representations that capture specifics of the target hardware, defined as quantum Instruction Set Architecture (ISA), like the device topology or the available basis gates. 

A large community effort has been devoted over the last decade to develop optimization approaches to the compilation of quantum circuits across their different abstraction levels to the target representation in ways that aim at being increasingly performant, both from an accuracy as well as a runtime point of view.  Tools like TKET~\cite{tket}, Cirq~\cite{cirq}, or, the most widely adopted, Qiskit, have enabled the quantum developer community with capabilities for circuit optimization and execution. 

When programming quantum computers, the quantum circuit can be further abstracted as part of a broader execution model. While the circuit remains central for expressing the computational complexity of the algorithm, this model includes additional elements that enhance or regulate the execution of the quantum program. This execution model can be represented as a directed acyclic graph which we will call a Tensor Compute Graph (TCG). In a TCG, nodes represent operations that input and output tensors, and edges connect output of operations to inputs of other operations. Whereas this is not a new concept in traditional HPC environments, framing it within a quantum context enhances the model in powerful ways towards problem scalability and algorithmic performance. In the quantum realm, examples of TCG nodes can be a quantum circuit; or a twirling graph, to generate random, structured circuit parameters; or some classical subroutine to post-process quantum circuit outputs. A quantum circuit node represents processing quantum information, relevant to specifics of a problem involving correlations, in ways that other constructs like tensor networks would fail to do at scale.

Within this execution model, quantum workflows that involve different quantum error mitigation or detection approaches can be expressed in a natural way. Furthermore, inter-dependencies across different components of hybrid algorithms become easier to express and orchestrate. Thus, different programming models, whether targeting tensors or quantum circuits, can coexist within this application middleware layer as defined by the applications layer described below. 

\subsection{QCSC Applications}
In the Applications layer of the architecture in Fig.~\ref{fig:arch}, the application box represents a set of libraries that offer an integrated, heterogeneous application‑development model. These libraries allow us to build scalable and performant domain‑specific solvers on top of them. To leverage quantum computing and further expand the utility of these solvers, the application libraries must be extended to support quantum embedding methods that support decomposing problems into mixed representations of tensors and quantum circuits. 

Quantum‑embedding capabilities transparently extend traditional embedding methods to encode sub-components of a domain problem for execution of a quantum computer. These encoding can be represented using domain‑relevant representations (e.g., fermionic or bosonic systems for Hamiltonian simulation, graphs for optimization). The domain‑specific solvers can then use quantum libraries to efficiently decompose the workflow for execution on a QCSC architecture.

In order to prepare an executable for the quantum sub-task in the application encoding, application libraries and solvers need to integrate quantum libraries that translate problem inputs, constraints, and tradeoffs into a quantum circuit representation. They provide application‑specific circuit‑synthesis techniques and post‑processing algorithms that enable performant and scalable integration within programmable workflows. Application‑specific circuit synthesis is a preparation step requiring modules for mapping domain representations to circuits, optimizing application constraints to encode tradeoffs, and developing strategies for system‑efficient encoding. The resulting circuit from this encoding and preparation pipeline then informs the selection of appropriate error‑mitigation and error‑correction methods, which produce the required TCG for execution on the target hardware. 

In addition to the preparation step, the application layer will also require quantum libraries for application-aware signal post-processing to accurately translate application results from probabilistic outputs of a quantum computer. SQD, as seen in several of the use cases earlier, is an example of such a post-processing algorithm. We expect this class of algorithms to grow as the community explores broader application experimentation using quantum computers.   

The methods discussed in this section are classical pre- and post-processing techniques that enable integrating quantum computation into an application workflow. Classical resource requirements for these subroutines vary widely depending on the method and problem encoding. This requires the libraries to be interoperable and optimized for different classical architectures.

\subsection{QCSC Cross Cutting Issues}
Integrating quantum systems with HPC systems presents cross-cutting challenges that span compute, system orchestration, middleware, and application layers. These challenges become increasingly critical as integration evolves from loosely coupled architectures to tightly coupled configurations requiring temporal and spatial co-location of quantum and classical systems.

\subsubsection{QCSC Cloud Software}
A key cross-cutting issue in the integration of quantum computing with HPC is the evolution of the cloud software stack that orchestrates services, containers, and resource management layers such as Kubernetes. This challenge is particularly significant because of the differences between traditional HPC scheduling models and emerging cloud-native orchestration systems. 

Today, most HPC systems are orchestrated through batch schedulers such as Slurm, LSF~\cite{ibm_lsf_hpc}, PBSPro~\cite{pbs}, which allocate large blocks of compute resources to long-running jobs and operate largely independent of cloud service architecture. In contrast, quantum computing platforms are typically exposed through cloud APIs and service layers, where quantum circuits are submitted remotely to a queue and executed asynchronously on quantum hardware. This architectural mismatch is generally manageable for loosely coupled workloads, where classical and quantum tasks interact infrequently and can tolerate queue delays and the relatively large network latency. 

However, this mismatch becomes increasingly problematic as the field moves towards tightly coupled hybrid workflows, where classical and quantum computations must interact iteratively and with low latency. As hybrid algorithms evolve and workloads demand temporal and spatial co-location of classical and quantum resources, the separation between HPC schedulers and cloud-native orchestration layers becomes a significant bottleneck. 

%\textbf{\textcolor{red}{Seelam: Would like to add a para about Quantum system API and executor primitive, i.e., quantum moving from batch model to a fine grained model}}

At the same time, broader trends in HPC are beginning to mitigate this challenge for a different reason. Traditionally, HPC systems were primarily used for large-scale modeling and simulation workloads, which are inherently batch-oriented. However, HPC infrastructure is increasingly being used for AI workloads, which include not only large-scale training but also fine-grained services such as real-time inference and web-based AI applications, there has been a growing adoption of cloud-native software stacks. Supporting these workloads has led to the growing adoption of cloud-native software stacks, containerized services, and orchestration frameworks such as Kubernetes and OpenShift within HPC environments~\cite{seelamvela}. 

This shift toward cloud-native infrastructure, driven largely by AI workloads, naturally aligns with the requirements of hybrid quantum-classical computing roadmap. As a result, the increasing use of container orchestration, microservices, and service-based architecture on AI-focused HPC systems is likely to provide a foundation for integrating quantum systems into these environments, enabling unified scheduling and resource management of heterogeneous resources across CPUs, GPUs, and QPUs. There is already work in this direction in the research community (see e.g., ~\cite{Qubernetes, kub-hpc, pilot-quantum}) and we will work to grow this work for QCSC systems.

\subsubsection{QCSC System Management and Monitoring}
%\subsubsection{QCSC System Management and Monitoring}
As quantum computing begins to integrate with HPC environments, system observability and management become critical cross-cutting concerns. Both HPC/AI systems and quantum computing platforms rely on complex hardware and software stacks that must be carefully managed to ensure reliability, performance, and provide metrics for efficient resource utilization. Quantum computers add unique dimensions to observability and management based on their scarcity and qubit fidelity, which potentially varies over space and time. 

Understanding the unique differences that quantum computing brings to the table is essential for identifying the challenges involved in integrating quantum systems into existing HPC infrastructures. In this section, we first summarize the system management and monitoring architectures used in modern HPC and AI systems, followed by a discussion on the same in quantum computing platforms. We then discuss the key developments required for these systems to converge as hybrid quantum-classical computing environment becoming more tightly integrated in the next few years. 
 
Modern HPC and AI systems rely on mature system management and monitoring stacks designed to operate at large scale across many nodes~\cite{seelamvela}. These stacks typically combine telemetry frameworks, and observability platforms to manage heterogeneous compute resources such as CPUs, GPUs, memory, storage, and high-speed interconnects.  Monitoring and observability are provided through layered telemetry systems that collect metrics across hardware and software stack with tools such as Prometheus~\cite{prometheus_monitoring}, Grafana~\cite{grafana_visualization} and node exporters~\cite{prometheus_node_exporter, nvidia_gpu_exporter, rocm_gpu_exporter}. These track resource utilization, network activity, power consumption, hardware health, etc. AI clusters with GPUs also incorporate accelerator-specific telemetry such as NVIDIA DCGM~\cite{nvidia_dcgm}, ROCm monitoring utilities~\cite{amd_rocm_smi}, and custom monitoring systems like IBM Autopilot~\cite{ibm_autopilot_github}, and Multinic CNI~\cite{ibm_multinic_cni} with monitoring for high bandwidth networks.

This mature ecosystem provides continuous observability and automated operational management for HPC and AI infrastructure enabling administrators to monitor large systems with thousands of CPUs, GPUs, complex networks, and many workloads efficiently as we described as example here for two AI systems with Grafana, Prometheus and node exporters~\cite{seelamvela}.

Quantum computing systems leverage some of these same technologies to monitor the classical compute portions of the system. For example, IBM is continuously monitoring our quantum systems using Prometheus metric collection, Open Telemetry, IBM Cloud Logs, and Grafana. Much like HPC environments, we have created alert thresholds and health indicators to be able to quickly understand bottlenecks in the system, and overall system health. Additional physical health characteristics of the quantum system includes monitoring physical systems, such as the cryogenic temperatures for the superconducting qubits.

One key difference for quantum computers is the fidelity of the qubits fluctuating over space and time. To achieve the best possible results from a quantum computing workload, the system must both be calibrated extremely accurately, using ideal pulses for each operation, and all system fidelities (qubit coherence, gate fidelities, readout errors) must be provided accurately to quantum compilers such as Qiskit SDK to make ideal qubit selection decisions based on the current state of the quantum computer. IBM Quantum systems periodically run automated monitoring experiments to detect drift from ideal calibration, and make fine grained adjustment to device tuning parameters if needed. In the event of larger quality drifts the systems alert for manual intervention. 

Another key difference is the scarcity of the QPU resource itself. When coupled in a classical HPC environment, the QPU remains the scarce resource. Ensuring efficient saturation of the QPU resource is key to extracting the most value out of a QCSC environment. This requires understanding what degree of classical pre- and post-processing must be done in the quantum runtime, to create an ideal level of parallelism in the environment. The saturation level of the QPU is very dependent on quantum workload, and can vary greatly based on the different kinds of workload running.

As hybrid quantum-classical computing evolves, the system management and monitoring infrastructure of HPC and quantum systems will need to converge to support increasingly tightly integrated workloads on future generation quantum systems. Hybrid algorithms require coordinated execution across classical and quantum resources, which in turn demand observability across the entire computational stack. 

The convergence of quantum and classical computing requires three key developments in monitoring and observability infrastructure.

First, quantum hardware telemetry must be integrated into standard HPC observability frameworks. Similar to how GPU metrics are exposed through tools like NVIDIA's DCGM or AMD's ROCm SMI, quantum systems need specialized node exporters that expose QPU-specific metrics—including qubit coherence times, gate errors, calibration status, and queue states—to existing monitoring platforms like Prometheus and Grafana. This enables unified monitoring across heterogeneous environments spanning CPUs, GPUs, and QPUs.

Second, monitoring infrastructure must provide cross-layer visibility spanning classical compute nodes, container orchestration, quantum runtime environments, and underlying quantum hardware. Drawing from GPU integration experiences where vendors developed specialized telemetry frameworks, quantum systems require monitoring layers that expose device-level telemetry while correlating with higher-level orchestration components including schedulers, container runtimes, and hybrid workflows. This cross-layer observability is essential for diagnosing performance issues in tightly coupled workloads.

Finally, as hybrid workloads achieve shot-level coupling between classical and quantum systems, coordinated performance analysis becomes critical. Following the model of GPU profiling tools like Nsight Systems~\cite{nsightsystems}, Nsight Compute~\cite{nsightcompute}, CUPTI~\cite{cupti}, ROCProf~\cite{rocprof}, and Omniperf~\cite{omniperf}, quantum computing needs a comparable ecosystem of profiling and observability tools that expose device telemetry, runtime behavior, and circuit execution characteristics, enabling developers to analyze and optimize interactions across the full hybrid computing stack.

\subsubsection{QCSC Security Architecture}

At the heart of QCSC is a classical distributed system. QCSC will, thus, support and incorporate existing best of breed practices and standards for systems, cloud, and cybersecurity. This means encryption is used for all in-flight and at rest data and all access points are appropriately firewalled and protected with mandatory access controls providing authentication and authorization. 

QCSC also poses unique security challenges that are absent from pure classical systems. These challenges stem from (i) dealing with heterogeneous programming and data models and API interfaces between classical and quantum computers, and (ii)
%Unique Quantum security challenges: classifying the quantum circuits to determine their risks and defining the isolation boundary of a Quantum task, i.e., Quantum System's unit of execution.
%For example, due to the different nature of quantum programs, the way to detect malicious or unsanctioned programs must be redesigned.
Quantum program pathways being inherently delay sensitive.
%; consequently, any security controls must obey the timing constraints.
Below, we discuss the security threats to QCSC and how to mitigate and manage those threats.

\nip[1ex]{Threat model.} 
%To define the threats, we first establish who are the actors and what needs to be protected. 
There are two parties in the simplest QCSC setting: the QCSC provider and its users. 
%In a QCSC setting, there are three parties that we must contend with. 
%The first is the user of QCSC,the second is the service/data provider (e.g., AI model service provider, VDB provider, etc.) and the third is the QCSC owner/maintainer/operator. 
The QCSC provider wants to protect the confidentiality, integrity, and availability of its software and hardware and that only authorized users can access the resources provided according to its policy. 
Users want to protect the confidentiality and integrity of their jobs, including the code, data, and the result of the computation. 

\nip[1ex]{Confidential Data and Execution Abstraction.}
%The compartmentalization and interpositioning mechanisms discussed above are necessary but insufficient when dealing with the overall issue of data access control, especially ensuring the confidentiality of data belonging to mutually distrusting parties.
Achieving technical guarantees of confidentiality requires a verified hardware root of trust to establish a Trusted Execution Environment (TEE).
However, configuring a TEE and ensuring a complex security policy is enforced across a distributed and heterogeneous environment, such as the one found in QCSC, can be time consuming and error-prone.
%Furthermore, the configuration cannot be static since the nature of data access and sharing will change based on the needs of a job.
%Hence, to satisfy a job-specific data confidentiality needs in a distributed and heterogeneous system like QCSC, 
We, thus, require system-wide automation along with a programming model extension to capture the data access policy.

We introduce a new programming and runtime abstraction called Confidential Code and Data Encapsulation (CDE).
CDE brings together a unified classical and Quantum computation model with secure data access control at the programming level. CDE is comprised of a programming data access model, a task encapsulation module, and a TEE-enhanced runtime for policy verification and enforcement. With a CDE model of execution, the user creates a job and encodes how and what data is accessed using a provided API. When a job is submitted to the QCSC, the workflow and jobs orchestrator extended with a CDE task encapsulation module, takes the job and decompose it into tasks and creates execution images that will get executed in a CDE runtime. The runtime is a confidential enclave with extra provisions to verify and control data accesses via a policy that is generated from combining the information encoded in the original submitted job and QCSC-specific policy.
The CDE execution abstraction can satisfy the need of a dynamic job-specific data protection and sharing model in a distributed and heterogeneous QCSC environment. 
%Data access verification and policy enforcements are done by leveraging the plugable interpositioning framework outlined above. 

As QCSC spans multiple classical and quantum systems, achieving CDE on QCSC requires two fundamental building blocks: (i) a mathematically verified compartmentalization mechanism in foundational operating software and firmware, and (ii) a unified system-wide security policy governance enforced via a granular inter-positioning framework.

\nip[1ex]{Provable Compartmentalization}. Compartmentalization of tasks ensure the tasks cannot access data or resources beyond what are assigned to them.
%It also prevents the failure of a task to affect other tasks in the system. 
To achieve security guarantees provided by CDE, it is critical to be able to compartmentalize tasks and the data they access in order to isolate different security domains.
%Compartmentalization must be done at the most privileged layer of the system to ensure the isolation mechanism cannot be compromised.
Typically, hardware-based isolation primitives are leveraged by a firmware or operating system to provide the technical means to perform compartmentalization.
%This firmware becomes a trusted component of the system and becomes part of the attack surface.
As this firmware is responsible for a fundamental security building block, it must be hardened and its attack surface reduced.

To provide the necessary hardening, the correctness of the implementation, the invariance that the compartmentalization provides and ultimately the security properties provided should be mathematically proven. 
To be practical, the core functionality of the firmware to enforce compartmentalization needs to be kept at a minimal.
The proofs are to be done layer by layer in a composable manner, with the next layer leveraging the proven properties of the layer underneath~\cite{ozga2025ace, Ozga2023}.
As the firmware can change over time due to specification changes or the discovery of bugs, this verification process must be done continuously. This can be achieved with the notion of verification-in-the-loop with formal specification and proofs embedded in the source code and proofs run within the continuous integration pipeline.
In this manner, all new code will get automatically verified before it gets deployed.

\nip[1ex]{Granular Interpositioning.} Enforcing security policies in CDE requires the ability to interpose on all accesses to data or resources at all levels of the platform.
These interposition locations, serving as policy enforcement points (PEP), include entry points for both external and internal APIs.
%Enforcement decisions are made by querying decision engines from the PEPs.

The security policies for QCSC need to cover a wide array of data and resource access control scenarios and at different granularity.
For example, a policy can determine the geo-location sensitivity of accessing certain data centers to the quantum circuits that can be co-located due to cross-talk concerns in the underlying quantum hardware.
%the amount of GPUs/QPUs needed by a task executing on a specific server.
To achieve maximum flexibility, the interpositioning framework is designed to be composable and extensible via security modules or plugins. 
These plugins not only provide specific policy enforcement implementation but also for observability tasks by logging and correlating accesses across the system.
The output of one plugin can be used by other plugins, thereby, allowing composition to create more sophisticated enforcement flows. 
By interposing on all communication pathways, the plugins can receive inputs from what is being interposed as well as consume information provided by other plugins and security modules to inform a decision.
These plugins are controlled and configured via a common security layer using a unified policy configuration and data model to provide consistent policy specification and enforcement. 
%The interposition implementation must take into account the exact location in the QCSC system where interposition is needed.
%Some locations do not have strict overhead requirements while other locations must be highly optimized, especially when dealing with quantum elements given the tight time requirements such accesses require.

%The security interposition must also occur between boundaries of a system. A single unified security policy can be applied across heterogeneous systems to provide consistent enforcement. In addition, given the heterogeneous nature of the platform, hardware rooted TEE environments are needed to isolate the protect confidentiality and integrity of user’s data. QCSC platform must provide the appropriate runtime encapsulation via software defined compartmentalization to ensure user’s computation and data are isolated.

\section{QCSC Reference Architecture Roadmap: A Phased Approach}
\label{sec:roadmap}
The reference architecture described in the previous section follows a phased implementation strategy as shown in Figure~\ref{fig:qcscroadmap}.  Each phase builds incrementally upon the previous one to progressively deepen the integration between quantum and classical computing systems. This evolutionary roadmap spans three distinct phases that reflect increasing levels of system integration and co-design maturity, while accounting for concurrent advancements in both classical systems technologies -- driven by AI and other emerging workloads --  and quantum systems technologies as we race to build fault-tolerant quantum computers~\cite{ibmquantumroadmap}.
    
\subsection{Phase 1: Quantum Systems as a co-processor to HPC}
 In the first phase, quantum systems are treated as highly specialized compute offload engines integrated within existing HPC infrastructures. This phase focuses on establishing foundational integration across multiple dimensions. 

     From a hardware integration perspective, quantum systems are initially deployed in close proximity to classical HPC systems, as demonstrated by first-of-kind installations at RIKEN~\cite{ibmriken2025systemtwo} and RPI~\cite{rpi_ibm_qsystemone_2024}. IBM Quantum System installations in datacenters require similar infrastructure to traditional classical equipment with a few specialized considerations.  One is the need for facilities that minimize vibration (often possible with on-grade and on-slab installations) and precautions that need to be taken to prevent electromagnetic interference equipment placement in close proximity to the system. Given our extensive experience in on-premise system installation to date and detailed specifications provided by IBM, it has been possible to integrate these requirements with existing buildings and infrastructure. In cases with demonstrated co-located Quantum and HPC infrastructure, the systems used proprietary high bandwidth networking so the network integration between these systems and IBM Quantum is done through an Ethernet shared network fabric, enabling basic data exchange and job submission capabilities. In summary, these early deployments demonstrate that successful on-prem quantum computing requires careful site preparation and infrastructure planning, to support reliable hybrid quantum-classical operation. Others have documented similar issues in the literature~\cite{laura-iqm}

 On the software side, significant progress continues to occur in extending resource management frameworks through standardized interfaces like QRMI and adapting scheduling systems such as Slurm to handle quantum resources, as we described in earlier section. These integrations enable quantum systems to be discovered, allocated, and managed within familiar HPC workflow paradigms, lowering the barrier to adoption for existing HPC users. We anticipate these integration efforts to accelerate and deepen as the systems mature and more user facilities start to bring quantum systems to their premises.

 As these hardware and software integration efforts mature, a growing ecosystem of applications emerges that leverage the integrated quantum-HPC systems and software to realized closed-loop hybrid computations. In addition, we anticipate an increased use of scale-up systems co-located the quantum system to develop novel error mitigation and early outer decoder algorithms to further improve the quality of quantum results.  These applications and algorithmic advances demonstrate the viability of quantum offloading for specific problem domains while informing the requirements for deeper integration in subsequent phases.

\subsection{Phase 2: Heterogeneous Quantum and Classical Systems}
The second phase marks a significant evolution towards tightly coupled heterogeneous architectures where quantum and classical resources operate as integrated computational platforms. This phase emphasizes reducing latency, enabling sophisticated feedback mechanisms, and supporting complex hybrid algorithms. These algorithms will have multiple feedback loops operating at different time granularities between quantum systems, scale-up systems (such as GPUs and specialized accelerators), and scale-out systems (such as distributed HPC/AI clusters) as shown in Figure~\ref{fig:arch}. These multi-tiered interactions enable workflows where fast inner loops handle real-time quantum-classical feedback, while slower outer loops coordinate broader computational tasks across distributed resources. 

From a hardware integration perspective, spatial and temporal co-location becomes increasingly critical. Quantum systems are no longer merely adjacent to classical infrastructure but are architecturally integrated with dedicated and co-designed scale-up infrastructure in close proximity to minimize communication latency. For this tighter coupling, specialized interconnects such as Ultra Ethernet~\cite{hoefler2025ultraethernetsdesignprinciples}, NVQLink~\cite{nvqlink}, and UALink~\cite{ualink_standard} may provide the ultra-low latency and bandwidth required for real-time feedback loops. In addition, these inter-operable interconnects may provide the foundation for the interconnecting scale-up systems with the scale-out systems for high-bandwidth, low-latency connections. We will strive to adapt and advance interoperable network fabrics that support standardized protocols to allow quantum and classical systems to communicate efficiently, regardless of vendor-specific implementations, facilitating seamless data movement and allow integrating Quantum systems with the broadest collection of HPC resources from different vendors. 

On the system management and middleware front, resource management frameworks evolve to handle unified scheduling and allocation across heterogeneous quantum-classical resources. Traditional HPC schedulers such as Slurm and LSF allocate requested resources including CPUs and GPUs but generally do not perform joint optimization of heterogeneous resources. Instead, they rely on user-defined resource requirements and queue-based scheduling policies such as priority, preemption, and backfilling. But quantum-classical heterogeneous systems require a fundamentally different approach. Advanced schedulers must now consider not only individual resource availability across  multiple resource types but also the complex dependencies and data flow patterns between quantum and classical components, including the timing constraints imposed by quantum coherence windows and the varying latency requirements of different feedback loops. Middleware layers become more sophisticated, providing abstractions for hybrid workflows, runtimes managing data movement between the three system boundaries, and orchestrating the complex execution graphs with different feedback loops.

\subsection{Phase 3: Tightly Integrated Quantum and HPC Systems}
This third phase represents the culmination of quantum-classical integration through fully co-designed heterogeneous systems where quantum and classical resources are architected as unified platforms from the ground up. 

This evolution mirrors the trajectory of GPUs in HPC systems. 
Early GPU-accelerated HPC systems connected GPUs to CPUs through PCIe bus, where GPUs functioned primarily as external accelerators attached to the host processor~\cite{arie04}. In this model, all data transfers between CPU and GPU and even between GPUs had to traverse the limited PCIe bandwidth. Systems such as Summit~\cite{summit} represent a significant evolution of this architecture.  A proprietary interconnect directly connects the GPUs to CPUs and  GPUs to GPUs, providing a much higher bandwidth, lower latency, and enabling tighter integration between heterogeneous processors. This shift transformed GPUs from loosely attached accelerators into tightly coupled compute components within the HPC systems. More recent AI systems extend this trend even further by designing CPUs specifically around the needs of GPUs in systems like NVIDIA Grace Blackwell systems~\cite{nvidia_grace_blackwell_2024}. In these systems, the CPU no longer simply a host orchestrating the tasks but instead forms a co-designed component of a unified heterogeneous computing system optimized for large-scale AI, simulation and modeling workloads.  

Similarly, quantum systems will transition from standalone units to fully integrated components within co-designed quantum-HPC platforms that optimize the interaction between quantum processors, classical accelerators, and distributed computing resources from the ground up.

We anticipate a unified programming model to emerge around this time to offer higher-level constructs that allow developers to express these complex algorithms more naturally, hiding much of the underlying complexity of heterogeneous execution and leveraging the underlying middleware runtime to handle it.

Furthermore, we expect quantum systems to support multi-tenant execution around this time which requires addressing the security challenges described previously. 

\section{Summary}
We have articulated our vision for a Quantum Centric Supercomputer and outlined a strategic roadmap for the evolution and co-design of our quantum systems with existing and future supercomputing infrastructure. This convergence will empower computational scientists to tackle some of the world's most challenging problems. 

We described five distinct use case exemplars that illustrate diverse temporal and spatial coupling requirements, including varying co-location, latency, and bandwidth constraints. Based on these exemplars, we proposed a staged system architecture that progressively integrates quantum systems with scale-up and scale-out classical compute infrastructure with an interoperable network and data fabric. Our architecture features a quantum resource management layer that enables existing and future workload management and scheduling layers to support hybrid quantum-classical workloads with the distinct requirements. Additionally, we described middleware and application layers that enable rich composition of these hybrid workloads while leveraging the extensive software ecosystem available on classical systems and growing ecosystem on quantum systems. 

We outlined how this architecture will be realized through three distinct phases, aligned with the maturity of our quantum systems, advancements in classical systems technology, and the evolution of use cases. We expect this reference architecture to evolve and mature as the community adopts it, builds systems, develops the software stack, and delivers new applications. 

We invite HPC and quantum application developers, software architects, HPC system designers, supercomputing centers, and fellow quantum system providers to engage with us in shaping the future of quantum-classical integration, contributing to the development of standards, best practices, and use cases to advance the field of quantum-centric supercomputing. 

\section{Acknowledgments}

We thank various members of the IBM Quantum team for providing valuable feedback on the reference architecture and draft versions of this paper, including Javier Robledo Moreno, Will Kirby, Kunal Sharma, Salvador de la Puente, Rajeev Malik, Rick Welp, and Stefan Wörner.

\bibliographystyle{IEEEtran}
\bibliography{main}

@article{kitaev1995quantum,
  title={Quantum measurements and the Abelian stabilizer problem},
  author={Kitaev, A Yu},
  journal={arXiv preprint quant-ph/9511026},
  year={1995}
}

@article{temme2017error,
  title={Error mitigation for short-depth quantum circuits},
  author={Temme, Kristan and Bravyi, Sergey and Gambetta, Jay M},
  journal={Physical review letters},
  volume={119},
  number={18},
  pages={180509},
  year={2017},
  publisher={APS}
}

@article{roncevic2026mobius,
author = {Igor Rončević  and Fabian Paschke  and Yueze Gao  and Leonard-Alexander Lieske  and Lene A. Gödde  and Stefano Barison  and Samuele Piccinelli  and Alberto Baiardi  and Ivano Tavernelli  and Jascha Repp  and Florian Albrecht  and Harry L. Anderson  and Leo Gross },
title = {A molecule with half-Möbius topology},
journal = {Science},
volume = {0},
number = {0},
pages = {eaea3321},
year = {2026},
doi = {10.1126/science.aea3321},
URL = {https://www.science.org/doi/abs/10.1126/science.aea3321},
eprint = {https://www.science.org/doi/pdf/10.1126/science.aea3321}
}

@article{Romero2025hubo,
   title={Bias-field digitized counterdiabatic quantum algorithm for higher-order binary optimization},
   volume={8},
   ISSN={2399-3650},
   url={http://dx.doi.org/10.1038/s42005-025-02270-3},
   DOI={10.1038/s42005-025-02270-3},
   number={1},
   journal={Communications Physics},
   publisher={Springer Science and Business Media LLC},
   author={Romero, Sebastián V. and Visuri, Anne-Maria and Cadavid, Alejandro Gomez and Simen, Anton and Solano, Enrique and Hegade, Narendra N.},
   year={2025},
   month=aug }

@misc{chandarana2025runtimeadvantage,
      title={Runtime Quantum Advantage with Digital Quantum Optimization}, 
      author={Pranav Chandarana and Alejandro Gomez Cadavid and Sebastián V. Romero and Anton Simen and Enrique Solano and Narendra N. Hegade},
      year={2025},
      eprint={2505.08663},
      archivePrefix={arXiv},
      primaryClass={quant-ph},
      url={https://arxiv.org/abs/2505.08663}, 
}

@article{glick2024kernels,
   title={Covariant quantum kernels for data with group structure},
   volume={20},
   ISSN={1745-2481},
   url={http://dx.doi.org/10.1038/s41567-023-02340-9},
   DOI={10.1038/s41567-023-02340-9},
   number={3},
   journal={Nature Physics},
   publisher={Springer Science and Business Media LLC},
   author={Glick, Jennifer R. and Gujarati, Tanvi P. and Córcoles, Antonio D. and Kim, Youngseok and Kandala, Abhinav and Gambetta, Jay M. and Temme, Kristan},
   year={2024},
   month=jan, pages={479–483} }

@misc{candelori2026learning,
      title={Shallow-circuit Supervised Learning on a Quantum Processor}, 
      author={Luca Candelori and Swarnadeep Majumder and Antonio Mezzacapo and Javier Robledo Moreno and Kharen Musaelian and Santhanam Nagarajan and Sunil Pinnamaneni and Kunal Sharma and Dario Villani},
      year={2026},
      eprint={2601.03235},
      archivePrefix={arXiv},
      primaryClass={quant-ph},
      url={https://arxiv.org/abs/2601.03235}, 
}

@article{Havlicek2019kernels,
   title={Supervised learning with quantum-enhanced feature spaces},
   volume={567},
   ISSN={1476-4687},
   url={http://dx.doi.org/10.1038/s41586-019-0980-2},
   DOI={10.1038/s41586-019-0980-2},
   number={7747},
   journal={Nature},
   publisher={Springer Science and Business Media LLC},
   author={Havlíček, Vojtěch and Córcoles, Antonio D. and Temme, Kristan and Harrow, Aram W. and Kandala, Abhinav and Chow, Jerry M. and Gambetta, Jay M.},
   year={2019},
   month=mar, pages={209–212} }

@misc{huang2025generative,
      title={Generative quantum advantage for classical and quantum problems}, 
      author={Hsin-Yuan Huang and Michael Broughton and Norhan Eassa and Hartmut Neven and Ryan Babbush and Jarrod R. McClean},
      year={2025},
      eprint={2509.09033},
      archivePrefix={arXiv},
      primaryClass={quant-ph},
      url={https://arxiv.org/abs/2509.09033}, 
}

@article{gao2022generative,
  title = {Enhancing Generative Models via Quantum Correlations},
  author = {Gao, Xun and Anschuetz, Eric R. and Wang, Sheng-Tao and Cirac, J. Ignacio and Lukin, Mikhail D.},
  journal = {Phys. Rev. X},
  volume = {12},
  issue = {2},
  pages = {021037},
  numpages = {26},
  year = {2022},
  month = {May},
  publisher = {American Physical Society},
  doi = {10.1103/PhysRevX.12.021037},
  url = {https://link.aps.org/doi/10.1103/PhysRevX.12.021037}
}

@misc{layden2025flowmodels,
      title={Wavefunction Flows: Efficient Quantum Simulation of Continuous Flow Models}, 
      author={David Layden and Ryan Sweke and Vojtěch Havlíček and Anirban Chowdhury and Kirill Neklyudov},
      year={2025},
      eprint={2510.08462},
      archivePrefix={arXiv},
      primaryClass={quant-ph},
      url={https://arxiv.org/abs/2510.08462}, 
}

@article{ameri2023vlasov,
   title={Quantum algorithm for the linear Vlasov equation with collisions},
   volume={107},
   ISSN={2469-9934},
   url={http://dx.doi.org/10.1103/PhysRevA.107.062412},
   DOI={10.1103/physreva.107.062412},
   number={6},
   journal={Physical Review A},
   publisher={American Physical Society (APS)},
   author={Ameri, Abtin and Ye, Erika and Cappellaro, Paola and Krovi, Hari and Loureiro, Nuno F.},
   year={2023},
   month=jun }

@misc{sachdeva2026integratedpipeline,
      title={Integrated error-suppressed pipeline for quantum optimization of nontrivial binary combinatorial optimization problems on gate-model hardware at the 156-qubit scale}, 
      author={Natasha Sachdeva and Gavin S. Hartnett and Smarak Maity and Samuel Marsh and Yulun Wang and Adam Winick and Ryan Dougherty and Daniel Canuto and You Quan Chong and G. Adam Cox and Michael Hush and Pranav S. Mundada and Christopher D. B. Bentley and Michael J. Biercuk and Yuval Baum},
      year={2026},
      eprint={2406.01743},
      archivePrefix={arXiv},
      primaryClass={quant-ph},
      url={https://arxiv.org/abs/2406.01743}, 
}

@article{bravyi2020variational,
  title = {Obstacles to Variational Quantum Optimization from Symmetry Protection},
  author = {Bravyi, Sergey and Kliesch, Alexander and Koenig, Robert and Tang, Eugene},
  journal = {Phys. Rev. Lett.},
  volume = {125},
  issue = {26},
  pages = {260505},
  numpages = {6},
  year = {2020},
  month = {Dec},
  publisher = {American Physical Society},
  doi = {10.1103/PhysRevLett.125.260505},
  url = {https://link.aps.org/doi/10.1103/PhysRevLett.125.260505}
}

@article{egger2021warmstart,
   title={Warm-starting quantum optimization},
   volume={5},
   ISSN={2521-327X},
   url={http://dx.doi.org/10.22331/q-2021-06-17-479},
   DOI={10.22331/q-2021-06-17-479},
   journal={Quantum},
   publisher={Verein zur Forderung des Open Access Publizierens in den Quantenwissenschaften},
   author={Egger, Daniel J. and Mareček, Jakub and Woerner, Stefan},
   year={2021},
   month=jun, pages={479} }

@misc{koch2025qoblib,
title={Quantum Optimization Benchmarking Library - The Intractable Decathlon}, 
author={Thorsten Koch and David E. Bernal Neira and Ying Chen and Giorgio Cortiana and Daniel J. Egger and Raoul Heese and Narendra N. Hegade and Alejandro Gomez Cadavid and Rhea Huang and Toshinari Itoko and Thomas Kleinert and Pedro Maciel Xavier and Naeimeh Mohseni and Jhon A. Montanez-Barrera and Koji Nakano and Giacomo Nannicini and Corey O'Meara and Justin Pauckert and Manuel Proissl and Anurag Ramesh and Maximilian Schicker and Noriaki Shimada and Mitsuharu Takeori and Victor Valls and David Van Bulck and Stefan Woerner and Christa Zoufal},
year={2025},
eprint={2504.03832},
archivePrefix={arXiv},
primaryClass={quant-ph},
url={https://arxiv.org/abs/2504.03832}, 
}

@article{kotil2025moo,
   title={Quantum approximate multi-objective optimization},
   volume={5},
   ISSN={2662-8457},
   url={http://dx.doi.org/10.1038/s43588-025-00873-y},
   DOI={10.1038/s43588-025-00873-y},
   number={12},
   journal={Nature Computational Science},
   publisher={Springer Science and Business Media LLC},
   author={Kotil, Ayse and Pelofske, Elijah and Riedmüller, Stephanie and Egger, Daniel J. and Eidenbenz, Stephan and Koch, Thorsten and Woerner, Stefan},
   year={2025},
   month=oct, pages={1168–1177} }

@article{farhi2014qaoa,
      title={A Quantum Approximate Optimization Algorithm}, 
      author={Edward Farhi and Jeffrey Goldstone and Sam Gutmann},
      year={2014},
      journal={arXiv preprint quant-ph/1411.4028}
}

@article{filippov2023scalable,
  title={Scalable tensor-network error mitigation for near-term quantum computing},
  author={Filippov, Sergei and Leahy, Matea and Rossi, Matteo AC and Garc{\'\i}a-P{\'e}rez, Guillermo},
  journal={arXiv preprint arXiv:2307.11740},
  year={2023}
}

@article{fischer2026dynamical,
  title={Dynamical simulations of many-body quantum chaos on a quantum computer},
  author={Fischer, Laurin E and Leahy, Matea and Eddins, Andrew and Keenan, Nathan and Ferracin, Davide and Rossi, Matteo AC and Kim, Youngseok and He, Andre and Pietracaprina, Francesca and Sokolov, Boris and others},
  journal={Nature Physics},
  pages={1--6},
  year={2026},
  publisher={Nature Publishing Group UK London}
}

@article{rall2019pauliprop,
  title = {Simulation of qubit quantum circuits via Pauli propagation},
  author = {Rall, Patrick and Liang, Daniel and Cook, Jeremy and Kretschmer, William},
  journal = {Phys. Rev. A},
  volume = {99},
  issue = {6},
  pages = {062337},
  numpages = {10},
  year = {2019},
  month = {Jun},
  publisher = {American Physical Society},
  doi = {10.1103/PhysRevA.99.062337},
  url = {https://link.aps.org/doi/10.1103/PhysRevA.99.062337}
}

@article{begusic2025cpt,
    author = {Begušić, Tomislav and Hejazi, Kasra and Chan, Garnet Kin-Lic},
    title = {Simulating quantum circuit expectation values by Clifford perturbation theory},
    journal = {The Journal of Chemical Physics},
    volume = {162},
    number = {15},
    pages = {154110},
    year = {2025},
    month = {04},
    issn = {0021-9606},
    doi = {10.1063/5.0269149},
    url = {https://doi.org/10.1063/5.0269149},
    eprint = {https://pubs.aip.org/aip/jcp/article-pdf/doi/10.1063/5.0269149/20490175/154110_1_5.0269149.pdf},
}

@misc{rudolph2025pauliprop,
      title={Pauli Propagation: A Computational Framework for Simulating Quantum Systems}, 
      author={Manuel S. Rudolph and Tyson Jones and Yanting Teng and Armando Angrisani and Zoë Holmes},
      year={2025},
      eprint={2505.21606},
      archivePrefix={arXiv},
      primaryClass={quant-ph},
      url={https://arxiv.org/abs/2505.21606}, 
}

@misc{fuller2025obp,
      title={Improved Quantum Computation using Operator Backpropagation}, 
      author={Bryce Fuller and Minh C. Tran and Danylo Lykov and Caleb Johnson and Max Rossmannek and Ken Xuan Wei and Andre He and Youngseok Kim and DinhDuy Vu and Kunal Sharma and Yuri Alexeev and Abhinav Kandala and Antonio Mezzacapo},
      year={2025},
      eprint={2502.01897},
      archivePrefix={arXiv},
      primaryClass={quant-ph},
      url={https://arxiv.org/abs/2502.01897}, 
}

@article{eddins2024lightcone,
  title={Lightcone shading for classically accelerated quantum error mitigation},
  author={Eddins, Andrew and Tran, Minh C and Rall, Patrick},
  journal={arXiv preprint arXiv:2409.04401},
  year={2024}
}

@article{zhuk2024trotter,
  title={Trotter error bounds and dynamic multi-product formulas for Hamiltonian simulation},
  author={Zhuk, Sergiy and Robertson, Niall F and Bravyi, Sergey},
  journal={Physical Review Research},
  volume={6},
  number={3},
  pages={033309},
  year={2024},
  publisher={APS}
}

@article{robertson2025tensor,
  title={Tensor network enhanced dynamic multiproduct formulas},
  author={Robertson, Niall F and Pokharel, Bibek and Fuller, Bryce and Switzer, Eric and Shtanko, Oles and Amico, Mirko and Byrne, Adam and D’Urbano, Andrea and Hayes-Shuptar, Salome and Akhriev, Albert and others},
  journal={PRX Quantum},
  volume={6},
  number={2},
  pages={020360},
  year={2025},
  publisher={APS}
}

@article{QuEPP2026,
  title = {Quantum Enhanced Pauli Propagation},
  author = {Majumder, Swarnadeep and Garrison, J. R. and Luo, L. and Mitchell, B. and Amico, M. and Seif, A. and Tran, M. and Sharma, K. and van den Berg, E. and Minev, Z. and Govia, C.},
  journal = {In preparation},
  year = {2026}
}

@article{PNA2026,
  title={Quantum Error Mitigation with Classically Propagated Noise Cancellation},
  author={Eddins, Andrew and others},
  journal={in prep},
  year={2026},
}

@article{li2017efficient,
  title={Efficient variational quantum simulator incorporating active error minimization},
  author={Li, Ying and Benjamin, Simon C},
  journal={Physical Review X},
  volume={7},
  number={2},
  pages={021050},
  year={2017},
  publisher={APS}
}

@article{lanes2025framework,
  title={A framework for quantum advantage},
  author={Lanes, Olivia and Beji, Mourad and Corcoles, Antonio D and Dalyac, Constantin and Gambetta, Jay M and Henriet, Loic and Javadi-Abhari, Ali and Kandala, Abhinav and Mezzacapo, Antonio and Porter, Christopher and others},
  journal={arXiv preprint arXiv:2506.20658},
  year={2025}
}

@article{yu2025quantum,
  title={Quantum-Centric Algorithm for Sample-Based Krylov Diagonalization},
  author={Yu, Jeffery and Moreno, Javier Robledo and Iosue, Joseph T and Bertels, Luke and Claudino, Daniel and Fuller, Bryce and Groszkowski, Peter and Humble, Travis S and Jurcevic, Petar and Kirby, William and others},
  journal={arXiv preprint arXiv:2501.09702},
  year={2025}
}

@misc{smith2025quantumcentricsimulationhydrogenabstraction_SQD,
      title={Quantum-centric simulation of hydrogen abstraction by sample-based quantum diagonalization and entanglement forging}, 
      author={Tyler Smith and Tanvi P. Gujarati and Mario Motta and Ben Link and Ieva Liepuoniute and Triet Friedhoff and Hiromichi Nishimura and Nam Nguyen and Kristen S. Williams and Javier Robledo Moreno and Caleb Johnson and Kevin J. Sung and Abdullah Ash Saki and Marna Kagele},
      year={2025},
      eprint={2508.08229},
      archivePrefix={arXiv},
      primaryClass={quant-ph},
      url={https://arxiv.org/abs/2508.08229}, 
}

@Article{Liepuoniute2025_SQD,
author={Liepuoniute, Ieva
and Doney, Kirstin D.
and Robledo Moreno, Javier
and Job, Joshua A.
and Friend, William S.
and Jones, Gavin O.},
title={Quantum-Centric Computational Study of Methylene Singlet and Triplet States},
journal={Journal of Chemical Theory and Computation},
year={2025},
month={May},
day={27},
publisher={American Chemical Society},
volume={21},
number={10},
pages={5062-5070},
issn={1549-9618},
doi={10.1021/acs.jctc.5c00075},
url={https://doi.org/10.1021/acs.jctc.5c00075}
}

@misc{shajan2026proteins_SQD,
      title={Molecular Quantum Computations on a Protein}, 
      author={Akhil Shajan and Danil Kaliakin and Fangchun Liang and Thaddeus Pellegrini and Hakan Doga and Subhamoy Bhowmik and Susanta Das and Antonio Mezzacapo and Mario Motta and Kenneth M. Merz Jr},
      year={2026},
      eprint={2512.17130},
      archivePrefix={arXiv},
      primaryClass={quant-ph},
      url={https://arxiv.org/abs/2512.17130}, 
}

@article{Barison_2025_ext-SQD,
doi = {10.1088/2058-9565/adb781},
url = {https://doi.org/10.1088/2058-9565/adb781},
year = {2025},
month = {feb},
publisher = {IOP Publishing},
volume = {10},
number = {2},
pages = {025034},
author = {Barison, Stefano and Robledo Moreno, Javier and Motta, Mario},
title = {Quantum-centric computation of molecular excited states with extended sample-based quantum diagonalization},
journal = {Quantum Science and Technology}
}

@article{robledo2024chemistry,
author = {Javier Robledo-Moreno  and Mario Motta  and Holger Haas  and Ali Javadi-Abhari  and Petar Jurcevic  and William Kirby  and Simon Martiel  and Kunal Sharma  and Sandeep Sharma  and Tomonori Shirakawa  and Iskandar Sitdikov  and Rong-Yang Sun  and Kevin J. Sung  and Maika Takita  and Minh C. Tran  and Seiji Yunoki  and Antonio Mezzacapo },
title = {Chemistry beyond the scale of exact diagonalization on a quantum-centric supercomputer},
journal = {Science Advances},
volume = {11},
number = {25},
pages = {eadu9991},
year = {2025},
doi = {10.1126/sciadv.adu9991},
URL = {https://www.science.org/doi/abs/10.1126/sciadv.adu9991},
eprint = {https://www.science.org/doi/pdf/10.1126/sciadv.adu9991}}

@misc{sqd-diag,
  author = {Tomonori Shirakawa},
  title = {Libary for selected basis diagonalization},
  howpublished = {\url{https://github.com/r-ccs-cms/sbd}},
  year = {2025},
  note = {Accessed: 2025-10-06}
}

@misc{qiskit-slurm,
  author = {},
  title = {Quantum spank plugins for Slurm},
  howpublished = {\url{https://github.com/qiskit-community/spank-plugins}},
  year = {2025},
}

@misc{qiskit-prefect,
  author = {},
  title = {Prefect Qiskit},
  howpublished = {\url{https://github.com/qiskit-community/prefect-qiskit}},
  year = {2025},
}

@misc{tket,
  author = {},
  title = {TKET},
  howpublished = {https://github.com/Quantinuum/tket},
  year = {2021},
}

@misc{qiskit,
  author = {},
  title = {Qiskit},
  howpublished = {https://github.com/Qiskit/qiskit},
  year = {2017},
}

@misc{cirq,
  author = {},
  title = {Cirq},
  howpublished = {https://github.com/quantumlib/Cirq},
  year = {2017},
}

@article{kanno2023QSCI,
  title={Quantum-Selected Configuration Interaction: classical diagonalization of Hamiltonians in subspaces selected by quantum computers},
  author={Kanno, Keita and Kohda, Masaya and Imai, Ryosuke and Koh, Sho and Mitarai, Kosuke and Mizukami, Wataru and Nakagawa, Yuya O},
  journal={arXiv:2302.11320},
  year={2023},
  url={https://arxiv.org/abs/2302.11320}
}

@misc{seelamvela,
      title={The infrastructure powering IBM's Gen AI model development}, 
      author={Talia Gershon and Seetharami Seelam and Brian Belgodere and Milton Bonilla and Lan Hoang and Danny Barnett and I-Hsin Chung and Apoorve Mohan and Ming-Hung Chen and Lixiang Luo and Robert Walkup and Constantinos Evangelinos and Shweta Salaria and Marc Dombrowa and Yoonho Park and Apo Kayi and Liran Schour and Alim Alim and Ali Sydney and Pavlos Maniotis and Laurent Schares and Bernard Metzler and Bengi Karacali-Akyamac and Sophia Wen and Tatsuhiro Chiba and Sunyanan Choochotkaew and Takeshi Yoshimura and Claudia Misale and Tonia Elengikal and Kevin O Connor and Zhuoran Liu and Richard Molina and Lars Schneidenbach and James Caden and Christopher Laibinis and Carlos Fonseca and Vasily Tarasov and Swaminathan Sundararaman and Frank Schmuck and Scott Guthridge and Jeremy Cohn and Marc Eshel and Paul Muench and Runyu Liu and William Pointer and Drew Wyskida and Bob Krull and Ray Rose and Brent Wolfe and William Cornejo and John Walter and Colm Malone and Clifford Perucci and Frank Franco and Nigel Hinds and Bob Calio and Pavel Druyan and Robert Kilduff and John Kienle and Connor McStay and Andrew Figueroa and Matthew Connolly and Edie Fost and Gina Roma and Jake Fonseca and Ido Levy and Michele Payne and Ryan Schenkel and Amir Malki and Lion Schneider and Aniruddha Narkhede and Shekeba Moshref and Alexandra Kisin and Olga Dodin and Bill Rippon and Henry Wrieth and John Ganci and Johnny Colino and Donna Habeger-Rose and Rakesh Pandey and Aditya Gidh and Aditya Gaur and Dennis Patterson and Samsuddin Salmani and Rambilas Varma and Rumana Rumana and Shubham Sharma and Aditya Gaur and Mayank Mishra and Rameswar Panda and Aditya Prasad and Matt Stallone and Gaoyuan Zhang and Yikang Shen and David Cox and Ruchir Puri and Dakshi Agrawal and Drew Thorstensen and Joel Belog and Brent Tang and Saurabh Kumar Gupta and Amitabha Biswas and Anup Maheshwari and Eran Gampel and Jason Van Patten and Matthew Runion and Sai Kaki and Yigal Bogin and Brian Reitz and Steve Pritko and Shahan Najam and Surya Nambala and Radhika Chirra and Rick Welp and Frank DiMitri and Felipe Telles and Amilcar Arvelo and King Chu and Ed Seminaro and Andrew Schram and Felix Eickhoff and William Hanson and Eric Mckeever and Michael Light and Dinakaran Joseph and Piyush Chaudhary and Piyush Shivam and Puneet Chaudhary and Wesley Jones and Robert Guthrie and Chris Bostic and Rezaul Islam and Steve Duersch and Wayne Sawdon and John Lewars and Matthew Klos and Michael Spriggs and Bill McMillan and George Gao and Ashish Kamra and Gaurav Singh and Marc Curry and Tushar Katarki and Joe Talerico and Zenghui Shi and Sai Sindhur Malleni and Erwan Gallen},
      year={2025},
      eprint={2407.05467},
      archivePrefix={arXiv},
      primaryClass={cs.DC},
      url={https://arxiv.org/abs/2407.05467}, 
}

@misc{shirakawa2025closedloopcalculationselectronicstructure,
      title={Closed-loop calculations of electronic structure on a quantum processor and a classical supercomputer at full scale}, 
      author={Tomonori Shirakawa and Javier Robledo-Moreno and Toshinari Itoko and Vinay Tripathi and Kento Ueda and Yukio Kawashima and Lukas Broers and William Kirby and Himadri Pathak and Hanhee Paik and Miwako Tsuji and Yuetsu Kodama and Mitsuhisa Sato and Constantinos Evangelinos and Seetharami Seelam and Robert Walkup and Seiji Yunoki and Mario Motta and Petar Jurcevic and Hiroshi Horii and Antonio Mezzacapo},
      year={2025},
      eprint={2511.00224},
      archivePrefix={arXiv},
      primaryClass={quant-ph},
      url={https://arxiv.org/abs/2511.00224}, 
}

@misc{tierkreis,
  title        = {Tierkreis},
  organization       = {{Quantinuum}},
  year         = {2024},
  howpublished = {https://github.com/Quantinuum/tierkreis},
}

@misc{fugaku_supercomputer,
  title        = {Fugaku Supercomputer},
  organization       = {{Riken Center for Computational Science}},
  year         = {2021},
  url = {https://www.r-ccs.riken.jp/en/fugaku/},
  note         = {Accessed: 2026-01-09; Fugaku is a petascale supercomputer developed by RIKEN and Fujitsu, deployed at the RIKEN Center for Computational Science in Kobe, Japan. It was ranked the world’s fastest in 2020 and 2021 on the TOP500 list.} 
}

@article{piccinelli2025quantum,
  title={Quantum chemistry with provable convergence via randomized sample-based quantum diagonalization},
  author={Piccinelli, Samuele and Baiardi, Alberto and Rossmannek, Max and Vazquez, Almudena Carrera and Tacchino, Francesco and Mensa, Stefano and Altamura, Edoardo and Alavi, Ali and Motta, Mario and Robledo-Moreno, Javier and others},
  journal={arXiv preprint arXiv:2508.02578},
  year={2025}
}

@INPROCEEDINGS{arie04,
  author={Zhe Fan and Feng Qiu and Kaufman, A. and Yoakum-Stover, S.},
  booktitle={SC '04: Proceedings of the 2004 ACM/IEEE Conference on Supercomputing}, 
  title={GPU Cluster for High Performance Computing}, 
  year={2004},
  volume={},
  number={},
  pages={47-47},
  doi={10.1109/SC.2004.26}}

@manual{qcsc,
  title        = {What is quantum-centric supercomputing?},
  organization = {IBM},
  year         = {2026},
  url          = {https://www.ibm.com/think/topics/quantum-centric-supercomputing},
}

@misc{nvqlink,
      title={Platform Architecture for Tight Coupling of High-Performance Computing with Quantum Processors}, 
      author={Shane A. Caldwell and Moein Khazraee and Elena Agostini and Tom Lassiter and Corey Simpson and Omri Kahalon and Mrudula Kanuri and Jin-Sung Kim and Sam Stanwyck and Muyuan Li and Jan Olle and Christopher Chamberland and Ben Howe and Bruno Schmitt and Justin G. Lietz and Alex McCaskey and Jun Ye and Ang Li and Alicia B. Magann and Corey I. Ostrove and Kenneth Rudinger and Robin Blume-Kohout and Kevin Young and Nathan E. Miller and Yilun Xu and Gang Huang and Irfan Siddiqi and John Lange and Christopher Zimmer and Travis Humble},
      year={2025},
      eprint={2510.25213},
      archivePrefix={arXiv},
      primaryClass={quant-ph},
      url={https://arxiv.org/abs/2510.25213}, 
}

@techreport{ibta_roce_2025, 
    author = {{InfiniBand Trade Association}}, 
    title = {InfiniBand Architecture Specification Volume 1, Release 1.4 (including RoCE)}, 
    institution = {InfiniBand Trade Association}, 
    year = {2020}, 
    note = {See also newer updates at https://www.infinibandta.org/ibta-specification/}, 
    url = {https://www.infinibandta.org/ibta-specification/}, 
    type = {Specification}
}

@misc{hoefler2025ultraethernetsdesignprinciples,
      title={Ultra Ethernet's Design Principles and Architectural Innovations}, 
      author={Torsten Hoefler and Karen Schramm and Eric Spada and Keith Underwood and Cedell Alexander and Bob Alverson and Paul Bottorff and Adrian Caulfield and Mark Handley and Cathy Huang and Costin Raiciu and Abdul Kabbani and Eugene Opsasnick and Rong Pan and Adee Ran and Rip Sohan},
      year={2025},
      eprint={2508.08906},
      archivePrefix={arXiv},
      primaryClass={cs.NI},
      url={https://arxiv.org/abs/2508.08906}, 
}

@misc{martiel2025lowoverheaderrordetectionspacetime,
      title={Low-overhead error detection with spacetime codes}, 
      author={Simon Martiel and Ali Javadi-Abhari},
      year={2025},
      eprint={2504.15725},
      archivePrefix={arXiv},
      primaryClass={quant-ph},
      url={https://arxiv.org/abs/2504.15725}, 
}

@article{Bausch2024HighAccuracyDecoding,
  author  = {Bausch, J. and Senior, A. W. and Heras, F. J. H. and Edlich, T. and Davies, A. and Newman, M. and Jones, C. and Satzinger, K. and Niu, M. Y. and Blackwell, S. and Holland, G. and Kafri, D. and Atalaya, J. and Gidney, C. and Hassabis, D. and Boixo, S. and Neven, H. and Kohli, P.},
  title   = {Learning high-accuracy error decoding for quantum processors},
  journal = {Nature},
  volume  = {635},
  pages   = {834--840},
  year    = {2024},
  doi     = {10.1038/s41586-024-08148-8},
  url     = {https://doi.org/10.1038/s41586-024-08148-8}
}

@article{Alexeev_2024,
   title={Quantum-centric supercomputing for materials science: A perspective on challenges and future directions},
   volume={160},
   ISSN={0167-739X},
   url={http://dx.doi.org/10.1016/j.future.2024.04.060},
   DOI={10.1016/j.future.2024.04.060},
   journal={Future Generation Computer Systems},
   publisher={Elsevier BV},
   author={Alexeev, Yuri and Amsler, Maximilian and Barroca, Marco Antonio and Bassini, Sanzio and Battelle, Torey and Camps, Daan and Casanova, David and Choi, Young Jay and Chong, Frederic T. and Chung, Charles and Codella, Christopher and Córcoles, Antonio D. and others},
   year={2024},
   month=nov, pages={666–710} }

@article{Bravyi2021MeasurementMitigation,
  author = {Bravyi, Sergey and Sheldon, Sarah and Kandala, Abhinav and Magesan, Easwar and Gambetta, Jay M.},
  title = {Mitigating Measurement Errors in Multiqubit Experiments},
  journal = {Physical Review A},
  volume = {103},
  pages = {042605},
  year = {2021},
  doi = {10.1103/PhysRevA.103.042605},
  eprint = {2006.14044},
  archivePrefix = {arXiv},
  primaryClass = {quant-ph}
}

@misc{jouppi2017indatacenterperformanceanalysistensor,
      title={In-Datacenter Performance Analysis of a Tensor Processing Unit}, 
      author={Norman P. Jouppi and others},
      year={2017},
      eprint={1704.04760},
      archivePrefix={arXiv},
      primaryClass={cs.AR},
      url={https://arxiv.org/abs/1704.04760}, 
}

@article{Bravyi2024LDPC,
  title   = {High-threshold and low-overhead fault-tolerant quantum memory},
  author  = {Bravyi, Sergey and Cross, Andrew W. and Gambetta, Jay M. and Maslov, Dmitri and Rall, Patrick and Yoder, Theodore J.},
  journal = {Nature},
  volume  = {627},
  pages   = {778--782},
  year    = {2024},
  doi     = {10.1038/s41586-024-07107-7},
  url     = {https://doi.org/10.1038/s41586-024-07107-7}
}

@misc{Yoder2025TourDeGross,
  title        = {Tour de gross: A modular quantum computer based on bivariate bicycle codes},
  author       = {Yoder, Theodore J. and Schoute, Eddie and Rall, Patrick and Pritchett, Emily and Gambetta, Jay M. and Cross, Andrew W. and Carroll, Malcolm and Beverland, Michael E.},
  year         = {2025},
  eprint       = {2506.03094},
  archivePrefix= {arXiv},
  primaryClass = {quant-ph},
  doi          = {10.48550/arXiv.2506.03094},
  url          = {https://arxiv.org/abs/2506.03094}
}

@misc{kirby2026observationimprovedaccuracyclassical,
      title={Observation of Improved Accuracy over Classical Sparse Ground-State Solvers using a Quantum Computer}, 
      author={William Kirby and Bibek Pokharel and Javier Robledo Moreno and Kevin C. Smith and Sergey Bravyi and Abhinav Deshpande and Constantinos Evangelinos and Bryce Fuller and James R. Garrison and Ben Jaderberg and Caleb Johnson and Petar Jurcevic and Su-un Lee and Simon Martiel and Mario Motta and Seetharami Seelam and Oles Shtanko and Kevin J. Sung and Minh Tran and Vinay Tripathi and Kazuhiro Seki and Kazuya Shinjo and Han Xu and Lukas Broers and Tomonori Shirakawa and Seiji Yunoki and Kunal Sharma and Antonio Mezzacapo},
      year={2026},
      eprint={2603.03496},
      archivePrefix={arXiv},
      primaryClass={quant-ph},
      url={https://arxiv.org/abs/2603.03496}, 
}

@inproceedings{laura-iqm, 
author = {Mansfield, Eric and Seegerer, Stefan and Vesanen, Panu and Echavarria, Jorge and Farooqi, Muhammad Nufail and Mete, Burak and Schulz, Laura}, 
title = {First Practical Experiences Integrating Quantum Computers with HPC Resources: A Case Study With a 20-qubit Superconducting Quantum Computer}, 
year = {2025}, 
isbn = {9798400718717}, 
publisher = {Association for Computing Machinery}, 
address = {New York, NY, USA}, url = {https://doi.org/10.1145/3731599.3767551}, 
doi = {10.1145/3731599.3767551}, 
booktitle = {Proceedings of the SC '25 Workshops of the International Conference for High Performance Computing, Networking, Storage and Analysis}, 
pages = {1842–1850}, numpages = {9}, 
location = { }, 
series = {SC Workshops '25} 
}

@article{Qubernetes, 
author = {Stirbu, Vlad and Kinanen, Otso and Haghparast, Majid and Mikkonen, Tommi}, 
title = {Qubernetes: Towards a unified cloud-native execution platform for hybrid classic-quantum computing}, 
year = {2024}, 
issue_date = {Nov 2024}, 
publisher = {Butterworth-Heinemann}, 
address = {USA}, 
volume = {175}, 
number = {C}, 
issn = {0950-5849}, 
url = {https://doi.org/10.1016/j.infsof.2024.107529}, 
doi = {10.1016/j.infsof.2024.107529}, 
journal = {Inf. Softw. Technol.}, 
month = nov, 
numpages = {11} 
}

@article{kub-hpc,
	author = {{Ciangottini, Diego} and {Spiga, Daniele} and {Memon, Ahmed Shiraz} and {Manzi, Andrea} and {Filipcic, Andrej} and {Troja, Antonino} and {Fanzago, Federica} and {Bianchini, Giulio} and {Sgaravatto, Massimo} and {Prica, Teo} and {Boccali, Tommaso} and {Tedeschi, Tommaso}},
	title = {Unlocking the compute continuum: Scaling out from cloud to HPC and HTC resources},
	DOI= "10.1051/epjconf/202533701296",
	url= "https://doi.org/10.1051/epjconf/202533701296",
	journal = {EPJ Web Conf.},
	year = 2025,
	volume = 337,
	pages = "01296",
}

@misc{pilot-quantum,
      title={Pilot-Quantum: A Quantum-HPC Middleware for Resource, Workload and Task Management}, 
      author={Pradeep Mantha and Florian J. Kiwit and Nishant Saurabh and Shantenu Jha and Andre Luckow},
      year={2025},
      eprint={2412.18519},
      archivePrefix={arXiv},
      primaryClass={quant-ph},
      url={https://arxiv.org/abs/2412.18519}, 
}

@misc{ibm_autopilot_github,
  title        = {Autopilot: A Tool to Detect Infrastructure Issues on Cloud-Native AI Systems},
  author       = {{IBM Research}},
  year         = {2025},
  howpublished = {\url{https://github.com/IBM/autopilot}},
  note         = {GitHub repository, accessed 2026-03-08}
}

@misc{ibm_multinic_cni,
  title        = {Multi-NIC CNI: High-Performance Container Networking for AI and HPC Workloads},
  author       = {{IBM Research }},
  year         = {2023},
  howpublished = {\url{https://github.com/foundation-model-stack/multi-nic-cni}},
  note         = {GitHub repository, accessed 2026-03-08}
}

@misc{nvidia_dcgm,
  title        = {NVIDIA Data Center GPU Manager (DCGM)},
  author       = {{NVIDIA Corporation}},
  year         = {2024},
  howpublished = {\url{https://github.com/NVIDIA/DCGM}},
  note         = {GPU monitoring and management suite for datacenter environments}
}

@misc{amd_rocm_smi,
  title        = {AMD System Management Interface (AMD SMI)},
  author       = {{Advanced Micro Devices }},
  year         = {2024},
  howpublished = {\url{https://github.com/ROCm/amdsmi}},
  note         = {GPU monitoring and management tools for the ROCm platform, accessed 2026-03-08}
}

@misc{prometheus_monitoring,
  title        = {Prometheus: Monitoring System and Time Series Database},
  author       = {{Prometheus Authors}},
  year         = {2024},
  howpublished = {\url{https://github.com/prometheus/prometheus}},
  note         = {Open-source systems monitoring and alerting toolkit, accessed 2026-03-08}
}

@misc{grafana_visualization,
  title        = {Grafana: The Open and Composable Observability and Data Visualization Platform},
  author       = {{Grafana Labs}},
  year         = {2024},
  howpublished = {\url{https://github.com/grafana/grafana}},
  note         = {Open-source platform for monitoring dashboards and observability, accessed 2026-03-08}
}

@misc{prometheus_node_exporter,
  title        = {Node Exporter},
  author       = {{Prometheus Authors}},
  year         = {2024},
  howpublished = {\url{https://github.com/prometheus/node\_exporter}},
  note         = {Prometheus exporter for hardware and OS metrics, accessed 2026-03-08}
}

@misc{nvidia_gpu_exporter,
  title        = {NVIDIA DCGM Exporter},
  author       = {{NVIDIA Corporation}},
  year         = {2024},
  howpublished = {\url{https://github.com/NVIDIA/dcgm-exporter}},
  note         = {Prometheus exporter for NVIDIA GPU metrics via DCGM, accessed 2026-03-08}
}

@misc{rocm_gpu_exporter,
  title        = {ROCm Exporter},
  author       = {{Advanced Micro Devices }},
  year         = {2024},
  howpublished = {\url{https://github.com/ROCm/rocm\_smi\_exporter}},
  note         = {Prometheus exporter for AMD GPU metrics via ROCm SMI, accessed 2026-03-08}
}

@misc{nsightsystems,
  title        = {NVIDIA Nsight Systems},
  author       = {{NVIDIA Corporation}},
  year         = {2026},
  howpublished = {\url{https://developer.nvidia.com/nsight-systems}},
  note         = {System-wide performance analysis tool for CPU-GPU workloads. Accessed: 2026-03-08}
}

@misc{nsightcompute,
  title        = {NVIDIA Nsight Compute},
  author       = {{NVIDIA Corporation authors}},
  year         = {2026},
  howpublished = {\url{https://developer.nvidia.com/nsight-compute}},
  note         = {Interactive kernel profiler for CUDA applications. Accessed: 2026-03-08}
}

@misc{cupti,
  title        = {{CUDA Profiling Tools Interface (CUPTI)}},
  author       = {{NVIDIA Corporation}},
  year         = {2026},
  howpublished = {\url{https://docs.nvidia.com/cupti/}},
  note         = {Low-level performance monitoring and tracing interface for CUDA applications. Accessed: 2026-03-08}
}

@misc{rocprof,
  title        = {{rocprof: ROCm Profiling Tool}},
  author       = {{Advanced Micro Devices}},
  year         = {2026},
  howpublished = {\url{https://rocmdocs.amd.com/projects/rocprofiler/en/latest/}},
  note         = {Performance profiling tool for AMD ROCm GPU applications. Accessed: 2026-03-08}
}

@misc{omniperf,
  title        = {{Omniperf: AMD GPU Performance Analysis Tool}},
  author       = {{Advanced Micro Devices } },
  year         = {2026},
  howpublished = {\url{https://github.com/AMDResearch/omniperf}},
  note         = {Open-source GPU performance analysis tool for AMD accelerators. Accessed: 2026-03-08}
}

@misc{qrmi,
      title={Quantum resources in resource management systems}, 
      author={Utz Bacher and Mark Birmingham and Christopher D. Carothers and Andrew Damin and Carlos D. Gonzalez Calaza and Ashwin Kumar Karnad and Stefano Mensa and Matthieu Moreau and Aurelien Nober and Munetaka Ohtani and Max Rossmannek and Philippa Rubin and M. Emre Sahin and Oscar Wallis and Amir Shehata and Iskandar Sitdikov and Aleksander Wennersteen},
      year={2025},
      eprint={2506.10052},
      archivePrefix={arXiv},
      primaryClass={quant-ph},
      url={https://arxiv.org/abs/2506.10052}, 
}

@misc{ibmquantumroadmap,
  title        = {{IBM Quantum Roadmap: Development and Innovation Roadmap for Scalable Quantum Computing}},
  author       = {{IBM Quantum}},
  year         = {2025},
  howpublished = {\url{https://www.ibm.com/roadmaps/quantum/}},
  note         = {Outlines IBM’s strategy for scaling quantum processors, integrating quantum with HPC workflows, and achieving large-scale fault-tolerant quantum computing. Accessed: 2026-03-08}
}

@misc{ibmriken2025systemtwo,
  title        = {{IBM and RIKEN Unveil First IBM Quantum System Two Outside of the U.S.}},
  author       = {{RIKEN}},
  year         = {2025},
  month        = jun,
  howpublished = {\url{https://www.r-ccs.riken.jp/en/outreach/topics/20250624-1/}},
  note         = {Press release, June 23, 2025. Accessed: 2026-03-08}
}

@misc{rpi_ibm_qsystemone_2024,
  title        = {Rensselaer Polytechnic Institute and IBM Unveil the World's First IBM Quantum System One on a University Campus},
  author       = {{Rensselaer Polytechnic Institute}},
  year         = {2024},
  month        = apr,
  howpublished = {\url{https://dotcio.rpi.edu/unveiling-ibm-quantum-system-one}},
  note         = {Accessed: 2026-03-08}
}

@INPROCEEDINGS{summit,
  author={Vazhkudai, Sudharshan S. and de Supinski, Bronis R. and Bland, Arthur S. and Geist, Al and Sexton, James and Kahle, Jim and Zimmer, Christopher J. and Atchley, Scott and Oral, Sarp and Maxwell, Don E. and Larrea, Veronica G. Vergara and Bertsch, Adam and Goldstone, Robin and Joubert, Wayne and Chambreau, Chris and Appelhans, David and Blackmore, Robert and Casses, Ben and Chochia, George and Davison, Gene and Ezell, Matthew A. and Gooding, Tom and Gonsiorowski, Elsa and Grinberg, Leopold and Hanson, Bill and Hartner, Bill and Karlin, Ian and Leininger, Matthew L. and Leverman, Dustin and Marroquin, Chris and Moody, Adam and Ohmacht, Martin and Pankajakshan, Ramesh and Pizzano, Fernando and Rogers, James H. and Rosenburg, Bryan and Schmidt, Drew and Shankar, Mallikarjun and Wang, Feiyi and Watson, Py and Walkup, Bob and Weems, Lance D. and Yin, Junqi},
  booktitle={SC18: International Conference for High Performance Computing, Networking, Storage and Analysis}, 
  title={The Design, Deployment, and Evaluation of the CORAL Pre-Exascale Systems}, 
  year={2018},
  volume={},
  number={},
  pages={661-672},
  doi={10.1109/SC.2018.00055}
}

@misc{bravyi2025quantumsimulationnoisyclassical,
      title={Quantum simulation of a noisy classical nonlinear dynamics}, 
      author={Sergey Bravyi and Robert Manson-Sawko and Mykhaylo Zayats and Sergiy Zhuk},
      year={2025},
      eprint={2507.06198},
      archivePrefix={arXiv},
      primaryClass={quant-ph},
      url={https://arxiv.org/abs/2507.06198}, 
}

@misc{ozga2025ace,
  author = {Ozga, Wojciech and Hunt, Guerney D. H. and Le, Michael V. and G{\"a}her, Lennard and Shinnar, Avraham and Palmer, Elaine R. and Jamjoom, Hani and Dragone, Silvio},
  title  = {ACE: Confidential Computing for Embedded RISC-V Systems},
  year   = {2025},
  eprint = {2505.12995},
  archivePrefix = {arXiv},
  primaryClass  = {cs.CR},
  doi    = {10.48550/arXiv.2505.12995}
}

@inproceedings{Ozga2023,
  author    = {Ozga, Wojciech and Hunt, Guerney D. H. and Le, Michael V. and Palmer, Elaine R. and Shinnar, Avraham},
  title     = {Towards a Formally Verified Security Monitor for VM-based Confidential Computing},
  booktitle = {Proc. 12th International Workshop on Hardware and Architectural Support for Security and Privacy (HASP '23)},
  year      = {2023},
  pages     = {73--81},
  doi       = {10.1145/3623652.3623668}
}

@misc{pattison2021improvedquantumerrorcorrection,
      title={Improved quantum error correction using soft information}, 
      author={Christopher A. Pattison and Michael E. Beverland and Marcus P. da Silva and Nicolas Delfosse},
      year={2021},
      eprint={2107.13589},
      archivePrefix={arXiv},
      primaryClass={quant-ph},
      url={https://arxiv.org/abs/2107.13589}, 
}

@misc{dasu2026computingencodedlogicalqubits,
      title={Computing with many encoded logical qubits beyond break-even}, 
      author={Shival Dasu and others},
      year={2026},
      eprint={2602.22211},
      archivePrefix={arXiv},
      primaryClass={quant-ph},
      url={https://arxiv.org/abs/2602.22211}, 
}

@article{Liao_2025,
   title={Achieving Computational Gains with Quantum Error-Correction Primitives: Generation of Long-Range Entanglement Enhanced by Error Detection},
   volume={6},
   ISSN={2691-3399},
   url={http://dx.doi.org/10.1103/PRXQuantum.6.020331},
   DOI={10.1103/prxquantum.6.020331},
   number={2},
   journal={PRX Quantum},
   publisher={American Physical Society (APS)},
   author={Liao, Haoran and Hartnett, Gavin S. and Kakkar, Ashish and Tan, Adrian and Hush, Michael and Mundada, Pranav S. and Biercuk, Michael J. and Baum, Yuval},
   year={2025},
   month=may }

@misc{reichardt2025faulttolerantquantumcomputationneutral,
      title={Fault-tolerant quantum computation with a neutral atom processor}, 
      author={Ben W. Reichardt and others},
      year={2025},
      eprint={2411.11822},
      archivePrefix={arXiv},
      primaryClass={quant-ph},
      url={https://arxiv.org/abs/2411.11822}, 
}

@article{Gupta_2024,
   title={Encoding a magic state with beyond break-even fidelity},
   volume={625},
   ISSN={1476-4687},
   url={http://dx.doi.org/10.1038/s41586-023-06846-3},
   DOI={10.1038/s41586-023-06846-3},
   number={7994},
   journal={Nature},
   publisher={Springer Science and Business Media LLC},
   author={Gupta, Riddhi S. and Sundaresan, Neereja and Alexander, Thomas and Wood, Christopher J. and Merkel, Seth T. and Healy, Michael B. and Hillenbrand, Marius and Jochym-O’Connor, Tomas and Wootton, James R. and Yoder, Theodore J. and Cross, Andrew W. and Takita, Maika and Brown, Benjamin J.},
   year={2024},
   month=jan, pages={259–263} }

@misc{brown2023advancescompilationquantumhardware,
      title={Advances in compilation for quantum hardware -- A demonstration of magic state distillation and repeat-until-success protocols}, 
      author={Natalie C. Brown and John Peter Campora III and Cassandra Granade and Bettina Heim and Stefan Wernli and Ciaran Ryan-Anderson and Dominic Lucchetti and Adam Paetznick and Martin Roetteler and Krysta Svore and Alex Chernoguzov},
      year={2023},
      eprint={2310.12106},
      archivePrefix={arXiv},
      primaryClass={quant-ph},
      url={https://arxiv.org/abs/2310.12106}, 
}

@inproceedings{slurm, 
author = {Jette, Morris A. and Wickberg, Tim}, 
title = {Architecture of the Slurm Workload Manager}, 
year = {2023}, 
isbn = {978-3-031-43942-1}, 
publisher = {Springer-Verlag}, 
address = {Berlin, Heidelberg}, 
url = {https://doi.org/10.1007/978-3-031-43943-8_1}, 
doi = {10.1007/978-3-031-43943-8_1}, 
booktitle = {Job Scheduling Strategies for Parallel Processing: 26th Workshop, JSSPP 2023, St. Petersburg, FL, USA, May 19, 2023, Revised Selected Papers}, 
pages = {3–23}, 
numpages = {21}, 
location = {St. Petersburg, FL, USA} 
}

@misc{ibm_lsf_hpc,
  author       = {{IBM Corporation}},
  title        = {IBM Spectrum LSF Suites: High-Performance Computing Workload Management},
  year         = {2024},
  url          = {https://www.ibm.com/products/hpc-workload-management},
  note         = {Online; accessed 9 March 2026}
}

@inbook{pbs, 
author = {Nitzberg, Bill and Schopf, Jennifer M. and Jones, James Patton}, 
title = {PBS Pro: Grid computing and scheduling attributes}, 
year = {2004}, 
isbn = {1402075758}, 
publisher = {Kluwer Academic Publishers}, 
address = {USA}, booktitle = {Grid Resource Management: State of the Art and Future Trends}, 
pages = {183–190}, 
numpages = {8} 
}

@misc{ualink_standard,
  author       = {{UALink Consortium}},
  title        = {Ultra Accelerator Link (UALink): Open Standard for AI Accelerator Interconnect},
  year         = {2025},
  howpublished = {\url{https://ualinkconsortium.org}},
  note         = {Accessed: 2026-03-09}
}

@article{kim2023evidence,
  title={Evidence for the utility of quantum computing before fault tolerance},
  author={Kim, Youngseok and Eddins, Andrew and Anand, Sajhin and Tran, Ken Xuan and Wang, Chin-Wen and Wang, Shaimaa and Edmands, J and Miller, J and others},
  journal={Nature},
  volume={618},
  number={7965},
  pages={500--505},
  year={2023},
  publisher={Nature Publishing Group UK London},
  doi={10.1038/s41586-023-06096-3},
  url={https://www.nature.com/articles/s41586-023-06096-3}
}

@misc{top500web,
  author = {E. Strohmaier and J. J. Dongarra and H. D. Simon and M. Meuer},
  title = {{TOP500} Supercomputer Sites},
  year = {2024},
  howpublished = {\url{https://www.top500.org}},
  note = {Accessed: 2024-03-10}
}

@article{Temme_2017,
   title={Error Mitigation for Short-Depth Quantum Circuits},
   volume={119},
   ISSN={1079-7114},
   url={http://dx.doi.org/10.1103/PhysRevLett.119.180509},
   DOI={10.1103/physrevlett.119.180509},
   number={18},
   journal={Physical Review Letters},
   publisher={American Physical Society (APS)},
   author={Temme, Kristan and Bravyi, Sergey and Gambetta, Jay M.},
   year={2017},
   month=nov 
}

@article{Li_2017,
   title={Efficient Variational Quantum Simulator Incorporating Active Error Minimization},
   volume={7},
   ISSN={2160-3308},
   url={http://dx.doi.org/10.1103/PhysRevX.7.021050},
   DOI={10.1103/physrevx.7.021050},
   number={2},
   journal={Physical Review X},
   publisher={American Physical Society (APS)},
   author={Li, Ying and Benjamin, Simon C.},
   year={2017},
   month=jun 
}

@article{PhysRevA.52.R2493,
  title = {Scheme for reducing decoherence in quantum computer memory},
  author = {Shor, Peter W.},
  journal = {Phys. Rev. A},
  volume = {52},
  issue = {4},
  pages = {R2493--R2496},
  numpages = {0},
  year = {1995},
  month = {Oct},
  publisher = {American Physical Society},
  doi = {10.1103/PhysRevA.52.R2493},
  url = {https://link.aps.org/doi/10.1103/PhysRevA.52.R2493}
}

@article{PhysRevLett.77.793,
  title = {Error Correcting Codes in Quantum Theory},
  author = {Steane, A. M.},
  journal = {Phys. Rev. Lett.},
  volume = {77},
  issue = {5},
  pages = {793--797},
  numpages = {0},
  year = {1996},
  month = {Jul},
  publisher = {American Physical Society},
  doi = {10.1103/PhysRevLett.77.793},
  url = {https://link.aps.org/doi/10.1103/PhysRevLett.77.793}
}

@article{Terhal_2015,
   title={Quantum error correction for quantum memories},
   volume={87},
   ISSN={1539-0756},
   url={http://dx.doi.org/10.1103/RevModPhys.87.307},
   DOI={10.1103/revmodphys.87.307},
   number={2},
   journal={Reviews of Modern Physics},
   publisher={American Physical Society (APS)},
   author={Terhal, Barbara M.},
   year={2015},
   month=apr, pages={307–346} }

@misc{nvidia_grace_blackwell_2024,
  title        = {{NVIDIA Grace Blackwell Platform}},
  author       = {{NVIDIA Corporation}},
  year         = {2024},
  howpublished = {\url{https://www.nvidia.com/en-us/data-center/technologies/blackwell-architecture/}},
  note         = {Accessed: 2026-03-11},
  organization = {NVIDIA}
}

@article{10.5555/2685179.2685184, 
author = {Gottesman, Daniel}, 
title = {Fault-tolerant quantum computation with constant overhead}, year = {2014}, 
issue_date = {November 2014}, 
publisher = {Rinton Press, Incorporated}, 
address = {Paramus, NJ}, 
volume = {14}, 
number = {15–16}, 
issn = {1533-7146}, 
journal = {Quantum Info. Comput.}, 
month = nov, 
pages = {1338–1372}, 
numpages = {35} 
}

@article{PhysRevLett.129.050504,
  title = {Constant-Overhead Quantum Error Correction with Thin Planar Connectivity},
  author = {Tremblay, Maxime A. and Delfosse, Nicolas and Beverland, Michael E.},
  journal = {Phys. Rev. Lett.},
  volume = {129},
  issue = {5},
  pages = {050504},
  numpages = {6},
  year = {2022},
  month = {Jul},
  publisher = {American Physical Society},
  doi = {10.1103/PhysRevLett.129.050504},
  url = {https://link.aps.org/doi/10.1103/PhysRevLett.129.050504}
}

@inproceedings{10.1145/3519935.3520017,
author = {Panteleev, Pavel and Kalachev, Gleb},
title = {Asymptotically good Quantum and locally testable classical LDPC codes},
year = {2022},
isbn = {9781450392648},
publisher = {Association for Computing Machinery},
address = {New York, NY, USA},
url = {https://doi.org/10.1145/3519935.3520017},
doi = {10.1145/3519935.3520017},
booktitle = {Proceedings of the 54th Annual ACM SIGACT Symposium on Theory of Computing},
pages = {375–388},
numpages = {14},
location = {Rome, Italy},
series = {STOC 2022}
}

@ARTICLE{10103665,
  author={Leverrier, Anthony and Zémor, Gilles},
  journal={IEEE Transactions on Information Theory}, 
  title={Decoding Quantum Tanner Codes}, 
  year={2023},
  volume={69},
  number={8},
  pages={5100-5115},
  doi={10.1109/TIT.2023.3267945}
}

@misc{ruchir,
      title={AI+HW 2035: Shaping the Next Decade}, 
      author={Deming Chen and Jason Cong and Azalia Mirhoseini and Christos Kozyrakis and Subhasish Mitra and Jinjun Xiong and Cliff Young and Anima Anandkumar and Michael Littman and Aron Kirschen and Sophia Shao and Serge Leef and Naresh Shanbhag and Dejan Milojicic and Michael Schulte and Gert Cauwenberghs and Jerry M. Chow and Tri Dao and Kailash Gopalakrishnan and Richard Ho and Hoshik Kim and Kunle Olukotun and David Z. Pan and Mark Ren and Dan Roth and Aarti Singh and Yizhou Sun and Yusu Wang and Yann LeCun and Ruchir Puri},
      year={2026},
      eprint={2603.05225},
      archivePrefix={arXiv},
      primaryClass={cs.AI},
      url={https://arxiv.org/abs/2603.05225}, 
}

\end{document}